\documentclass[aps,twocolumn,superscriptaddress,floatfix,preprintnumbers,nofootinbib,showpacs]{revtex4-1}
\usepackage[utf8]{inputenc}

\pdfoutput=1

\usepackage[english]{babel}
\usepackage{amsmath,amssymb,amsfonts}
\usepackage{graphicx}
\usepackage{booktabs}
\usepackage{hyperref}
\usepackage{color,xcolor}
\usepackage[sort&compress]{natbib}
\usepackage{tikz}
\usepackage{cleveref}
\usepackage{ulem}
\usepackage{slashed}
\usepackage[makeroom]{cancel}
\usepackage{url}

\usepackage{comment}
\usepackage{longtable}

\def\beqn{\begin{eqnarray}}
\def\eeqn{\end{eqnarray}}
\def\beqs{\begin{subequations}}
\def\eeqs{\end{subequations}}
\def\beq{\begin{equation}}
\def\eeq{\end{equation}}
\def\ba{\begin{array}}
\def\ea{\end{array}}

\def\non{\nonumber\\}

\def\[{\left[}
\def\]{\right]}
\def\({\left(}
\def\){\right)}

\def\GeV{\rm GeV}

\def\gU{\rm U}
\def\gSU{\rm SU}

\def\mC{\mathcal{C}}

\def\mG{\mathcal{G}}

\def\mL{\mathcal{L}}

\def\mO{\mathcal{O}}

\def\mR{\mathcal{R}}

\newcommand{\Tr}   {~\mathrm{Tr}}

\begin{document}

\title{Axion model with the ${\rm SU}(6)$ unification}

\author{Ning Chen}
\email[]{ustc0204.chenning@gmail.com}
\author{Yutong Liu}
\email[]{yutongliu@mail.nankai.edu.cn}
\author{Zhaolong Teng}
\email[]{tengcl@mail.nankai.edu.cn}
\affiliation{School of Physics, Nankai University, Tianjin, 300071, China}
%

\begin{abstract} 
We propose an ${\rm SU}(6)$ unified theory that solves the strong {\it CP} problem with a high-quality axion.
This is achieved by an automatic global Peccei-Quinn $\gU(1)_{\rm PQ}$ symmetry and the gauged discrete $\mathbb{Z}_{4\mR}$ symmetry.
With the axion mass predictions of $m_a \sim \mO(10^{-3} ) - \mO(0.1) \,{\rm eV}$, as well as a universal axion-photon coupling in the unification models, the QCD axion can be probed in the upcoming experiments of IAXO.
An intermediate gauge symmetry breaking scale characterized by the axion decay constant is obtained by the Peccei-Quinn quality argument, which is likely to achieve a successful leptogenesis through the type-I seesaw mechanism.
The Higgs sector at the electroweak scale is a type-II two-Higgs-doublet model.
The gauge coupling unification is possible with the supersymmetric extension.
\end{abstract}

\maketitle

\section{Introduction}
\label{sec:intro}
%
%
Grand unified theories (GUTs)~\cite{Georgi:1974sy,Fritzsch:1974nn} were proposed to unify all fundamental interactions.
In the original Georgi-Glashow ${\rm SU}(5)$ GUT~\cite{Georgi:1974sy}, it was assumed that {\it there are no unobserved fermions beyond the Standard Model (SM)}.
Aside from the aesthetic aspect of achieving the gauge coupling unification in its supersymmetric (SUSY) version, it is pragmatic to envision that {\it a successful GUT could address as many physical issues beyond the SM as possible.}
One leading puzzle of the SM is the strong {\it CP} problem, which predicts the neutron electric dipole moment due to the nontrivial topological term of $\theta \frac{ \alpha_{3c} }{ 8\pi  } G_{\mu\nu}^a \tilde G^{\mu\nu\, a}$ in the quantum chromodynamics (QCD) sector. 
However, the experimental upper limit of $|d_n| < 3.0\cdot 10^{-26}\, e\, {\rm cm}$~\cite{Afach:2015sja} leads to an extremely tiny upper bound of $| \bar \theta | \lesssim 10^{-10} $, where $\bar \theta$ is shifted from the $\theta$ parameter by the phase of the quark mass matrix $M_q$ as $\bar \theta= \theta + {\rm arg } \det M_q$. 
Axion field, which transforms under a global ${\rm U}(1)$ Peccei-Quinn (PQ) symmetry~\cite{Peccei:1977hh,Weinberg:1977ma,Wilczek:1977pj}, is the most appealing candidate to solve the strong {\it CP} problem since its first proposal.
Two widely studied benchmark scenarios of the invisible axion models are the Kim-Shifman-Vainshtein-Zakharov model~\cite{Kim:1979if,Shifman:1979if} and the Dine-Fischler-Srednicki-Zhitnitsky model~\cite{Zhitnitsky:1980tq,Dine:1981rt}.

In the past four decades, the GUTs were considered as frameworks for the axion models~\cite{Wise:1981ry,Lazarides:1981kz}.
Most studies based on the ${\rm SU}(5)$~\cite{Georgi:1981pu,Rubakov:1997vp,Co:2016xti,Lee:2016wiy,Boucenna:2017fna,Daido:2018dmu,DiLuzio:2018gqe,Ernst:2018rod,FileviezPerez:2019fku,FileviezPerez:2019ssf}, ${\rm SO}(10)$~\cite{Bajc:2005zf,Altarelli:2013aqa,Babu:2015bna,Ernst:2018bib,Ernst:2018rod,Coriano:2019vjl,DiLuzio:2020qio}, and $E_6$~\cite{Coriano:2017ghp} groups assumed a global ${\rm U}(1)_{\rm PQ}$ symmetry that commutes with the GUT group.
This leads to the PQ quality problem~\cite{Georgi:1981pu,Dine:1986bg,Barr:1992qq,Kamionkowski:1992mf,Holman:1992us,Ghigna:1992iv}.
It is therefore desirable to have the global PQ symmetry arise automatically, and there have been discussions in models with extended gauge symmetries~\cite{Carpenter:2009zs,Harigaya:2013vja,Dias:2014osa,Harigaya:2015soa,Redi:2016esr,Higaki:2016yqk,Fukuda:2017ylt,DiLuzio:2017tjx,Duerr:2017amf,Lillard:2018fdt,Lee:2018yak,Gavela:2018paw,Ardu:2020qmo,Yin:2020dfn,Nakai:2021nyf}.

In this paper, we propose an ${\rm SU}(6)$ GUT with a high-quality ${\rm U}(1)_{\rm PQ}$.
It was first noted by Dimopoulos, Raby, and Susskind (DRS)~\cite{Dimopoulos:1980hn} that an anomaly-free ${\rm SU}(N+4)$ gauge theory with $N$ antifundamental chiral fermions and one rank-$2$ antisymmetric chiral fermion enjoys a global symmetry of 
\beqn\label{eq:DRS_global}
\mG_{\rm global}&=& {\rm SU}(N)_F \otimes {\rm U}(1)_{\rm PQ}\,,
\eeqn
when $N \geq 2$.
The ${\rm SU}(6)$ is thus the minimal GUT group with an automatic PQ symmetry.
Historically, the ${\rm SU}(6)$ GUT was studied in various contexts~\cite{Abud:1977ej,Langacker:1977ai,Inoue:1977qd,Inoue:1977an,Inoue:1977qw,Yun:1978aa,Langacker:1978fn,Tomozawa:1978uz,Georgi:1978bv,Kim:1981jw,Fukugita:1981gn,Tabata:1983cr,Sen:1983xj,Barr:2011cz,Deppisch:2016jzl,Li:2019qxy,Chacko:2020tbu,Angelescu:2021nbp}, and found to alleviate the intrinsic doublet-triplet splitting problem~\cite{Berezhiani:1989bd,Dvali:1993yf,Berezhiani:1995dt,Berezhiani:1995sb,Barr:1997pt,Chacko:1998zz} and the proton lifetime constraint~\cite{Langacker:1977ai,Langacker:1978fn,Sen:1983xj} in the Georgi-Glashow ${\rm SU}(5)$ GUT~\cite{Georgi:1974sy}.

The rest of the paper is organized as follows.
In Sec.~\ref{sec:mini}, we setup our minimal SUSY ${\rm SU}(6)$ model.
The physical ${\rm SU}(6)$ axion is derived in Sec.~\ref{sec:SU6axion}.
Our main observation is given in Sec.~\ref{sec:quality}, where a high-quality QCD axion can arise with axion decay constant of $f_a\sim \mO(10^8) - \mO(10^{10})\,\GeV$ in the SUSY-extended ${\rm SU}(6)$, and be probed in upcoming IAXO experiment.
An intermediate symmetry of $\mG_{331}={\rm SU}(3)_c\otimes {\rm SU}(3)_L \otimes {\rm U}(1)_N$~\cite{Pisano:1991ee,Foot:1992rh,Ng:1992st,Tonasse:1996cx,Dias:2002gg,Ferreira:2011hm,Li:2019qxy} is found to break at the scale of $v_{331}\sim \mO(10^9)- \mO(10^{11})\,\GeV$, which is obtained from the PQ-quality argument.
The possible axion domain wall is briefly discussed in Sec.~\ref{sec:DW}.
The gauge coupling unification is indicated in Sec.~\ref{sec:unification}.
We conclude and envision some future efforts in {\it nonminimal} GUTs in Sec.~\ref{sec:conclusion}.
In Appendix.~\ref{sec:proof}, we prove that the non-SUSY ${\rm SU}(6)$ model cannot lead to a high-quality axion.
The details of the ${\rm SU}(6)$ symmetry breaking and the spectrum can be found in Appendix.~\ref{sec:SU6spectrum}.

\section{The minimal model}
\label{sec:mini}
%
%
The minimal setup of an anomaly-free SUSY ${\rm SU}(6)$ model is made up of three generational chiral superfields of $\mathbf{\bar 6_F^{\rho=\mathrm{I} \,, \mathrm{II} } }$ and $\mathbf{15_F}$ that contain all necessary fermions.
Two $\mathbf{\bar 6_F^{\rho} }$ form an $\gSU(2)_F$ doublet, which is free from the Witten anomaly~\cite{Witten:1982fp}.
The ${\rm SU}(6)$ undergoes two stages of symmetry breaking~\cite{Li:1973mq,Chen:2010er,Deppisch:2016jzl,Li:2019qxy}
\beqn\label{eq:pattern}
&& {\rm SU}(6) \xrightarrow{\Lambda_{\rm GUT} } {\cal G}_{331}  \xrightarrow{ v_{331}} {\cal G}_{\rm SM}  \,, \non
&& \mG_{\rm SM} =  {\rm SU}(3)_c \otimes {\rm SU}(2)_L \otimes {\rm U}(1)_Y\,,
\eeqn
in our discussions.

Next, we determine the Higgs sector of the minimal $\gSU(6)$ model by imposing the requirements that
\begin{enumerate}

    \item A superfield of $\mathbf{35_H}$~\cite{Li:1973mq} with its scalar vacuum expectation value (VEV) at $\Lambda_{\rm GUT}$ is necessary to achieve the first-stage symmetry breaking in Eq.~\eqref{eq:pattern}.

    \item All Yukawa couplings of the $125\,\GeV$ SM-like Higgs boson should be reproduced.
    This leads to at least one $\mathbf{\bar 6_{H\, \mathrm{I} } }$ and one $\mathbf{15_H}$.

    \item  Two $\mathbf{\bar 6_{H\,\rho} }$ are necessary to keep the ${\rm SU}(2)_F$-invariant couplings with two antifundamental superfields of $\mathbf{\bar 6_F^\rho}$.
    Furthermore, only one of them (which we chose to be $\mathbf{\bar 6_{H\,\mathrm{II} } }$ without loss of generality) are allowed to develop VEV and give fermion masses at $v_{331}$.
    Otherwise the low-energy effective theory is not anomaly free.

    \item A superfield of $\mathbf{21_H}$~\cite{Li:2019qxy}, together with the $\mathbf{\bar 6_{H\, \mathrm{II}}}$, achieve the second-stage symmetry breaking and the PQ-symmetry breaking through their scalar VEVs.
     Another conjugate superfield of $\mathbf{\overline{21}_H }$ is necessary for the anomaly cancellation.  
    
\end{enumerate}

\begin{table}[htp]
\begin{center}
\begin{tabular}{c|cc|ccccc}
\hline
 &  $ \mathbf{ \bar 6_F^\rho} $ & $\mathbf{15_F}$ & $ \mathbf{\bar 6_{H\,\rho}} $ & $\mathbf{15_H}$ & $\mathbf{21_H}$ &  $\mathbf{\overline{21}_H}$  & $ \mathbf{35_H} $   \\
\hline
${\rm SU}(2)_F$  &  $\Box$  & $1$  &  $\overline \Box$  & $1$  & $1$  & $1$ & $1$  \\
${\rm U}(1)_{\rm PQ}$  & $1$ & $1$ & $-2$ & $- 2$ & $-2 $ &  $0$  &  $0$  \\
$\mathbb{Z}_{4\mR}$  &  $0$  & $0$  & $2$ & $2$  & $2$  & $0$ & $0$   \\
\hline
\end{tabular}
\end{center}
\caption{The superfields with their ${\rm SU}(2)_F$ representations, and the charges under the ${\rm U}(1)_{\rm PQ}$ and $\mathbb{Z}_{4\mR}$ symmetries, in the SUSY ${\rm SU}(6)$ model.}
\label{tab:SU6min_PQ}
\end{table}%

Collectively, we tabulate the chiral superfields with their PQ and $\mathbb{Z}_{4\mR}$ charges in Table~\ref{tab:SU6min_PQ}.
A discrete and gauged $\mathbb{Z}_{4\mR}$ symmetry with proper charge assignments is necessary to avoid the dangerous dimension-three PQ-breaking operators.
The ${\rm SU}(2)_F$ is also gauged in order to avoid the constraint from the gravity.
Indeed, the mixed gauge anomalies cancel as follows
\beqs
\beqn
\mathbb{Z}_{4\mR} [{\rm SU}(6)]^2&=& n_g \times [ 2\times (-1) + 4(-1) ]  \non
&+& 2 \times 1 + 4\times 1 + 8 \times 1 \non
&+& 12 = 0 ~ {\rm mod}~ 4 \,, \\
\mathbb{Z}_{4\mR} [{\rm SU}(2)_F ]^2&=& n_g\times 6 \times (-1) + 6 \times 1 + 4 \non
&=& 0 ~ {\rm mod}~ 4  \,,
\eeqn
\eeqs
with $n_g=3$.
Accordingly, the superpotential includes the following terms
\beqn\label{eq:SU6W}
W_Y&=&  \mathbf{15_F}  \mathbf{\bar 6_F^\rho } \mathbf{\bar 6_{H\,\rho} } + \mathbf{ 15_F} \mathbf{ 15_F} \mathbf{ 15_H} \non
&+& \epsilon_{\rho\delta} \mathbf{\bar 6_F^\rho }  \mathbf{\bar 6_F^\delta }   \mathbf{ 15_H} + \epsilon_{\rho\delta}  \mathbf{\bar 6_F^\rho} \mathbf{\bar 6_F^\delta}   \mathbf{ 21_H} \,.
\eeqn
Schematically, we denote the Higgs VEVs and their hierarchies as follows
\beqn\label{eq:VEV_hierarchy}
&&  \langle \mathbf{35_H} \rangle \sim \Lambda_{\rm GUT} \,,\non
&& \langle   \mathbf{\bar 6_{H\,\mathrm{II}} } \rangle = v_3\,, \quad \langle \mathbf{21_H} \rangle = v_6\,,\quad v_3 \sim v_6\sim v_{331} \,, \non
&& \langle \mathbf{\bar 6_{H\,\mathrm{I}} } \rangle = v_d = v_{\rm EW} \sin\beta \,,\quad  \langle \mathbf{15_H} \rangle = v_u = v_{\rm EW} \cos\beta\,,   \non
&& \Lambda_{\rm GUT} \gg v_{331}  \gg v_{\rm EW} =( \sqrt{2} G_F)^{-1/2} \simeq 246\,\GeV\,,
\eeqn
with $\tan\beta$ being the ratio between two EW Higgs VEVs.

After the first-stage symmetry breaking in Eq.~\eqref{eq:pattern}, the $( \mathbf{ 1}\,, \mathbf{\bar 3}\,, -\frac{1}{3} )_{\mathbf{H}\, {\rm II}} \subset  \mathbf{\bar 6_{H\,\mathrm{II}}} $ and the $ ( \mathbf{ 1}\,, \mathbf{ 6}\,, +\frac{2}{3} )_{\mathbf{H}} \subset  \mathbf{ 21_H }$ will develop their VEVs to trigger the second-stage symmetry breaking.
The Yukawa couplings of $\mathbf{15_F} \mathbf{\bar 6_F^{\mathrm{II } }} \mathbf{\bar 6_{H\,\mathrm{II}} } + {\rm H.c.}$ give masses to the vectorlike $D$ quarks of $m_D \simeq \mO(v_{331} )$.
With the electric charge of $-1/3$, the $D$ quarks can form $d=4$ operators from the Yukawa couplings in Eq.~\eqref{eq:SU6W}, and the $D$-hadron lifetime~\cite{DiLuzio:2015oha,DiLuzio:2016sbl,DiLuzio:2017pfr} is found to be $\tau_D \sim m_D^{-1} \sim \mO(10^{-36})- \mO(10^{-34}) \,{\rm sec}$ with $v_{331} \sim \mO(10^{9} ) - \mO(10^{11} ) \,\GeV$ through the PQ quality analysis below.
This satisfies the cosmological constraint of $\tau_Q \lesssim 10^{-2}\, {\rm sec}$~\cite{Kawasaki:2004qu,Jedamzik:2007qk} from the big bang nucleosynthesis.
In addition, the Yukawa couplings of $ \mathbf{\bar 6_F^{[ \mathrm{I } }}  \mathbf{\bar 6_F^{\mathrm{II } ] }}  \mathbf{21_H } + {\rm H.c.}$ gives heavy neutrino masses of $m_{N\,, N^\prime}\simeq \mO(v_{331})$, which satisfy the Davidson-Ibarra bound~\cite{Davidson:2002qv,Buchmuller:2002rq,Ellis:2002xg} of $M_N \gtrsim 10^9\,{\rm GeV}$ for a successful leptogenesis.
Two ${\rm SU}(2)_L$ EW Higgs doublets come from the $\mathbf{\bar 6_H^{\mathrm{I}} }$ and the $\mathbf{15_H}$ and lead to the type-II 2HDM.
Besides, a type-I seesaw mechanism can also be realized with the Yukawa couplings of $\mathbf{\bar 6_F^{[ \mathrm{I } }}  \mathbf{\bar 6_F^{\mathrm{II } ] }}  \mathbf{15_H } + {\rm H.c.}$.
All relevant Yukawa couplings in two stages of symmetry breaking will be explicitly given in Eqs.~\eqref{eqs:331Yukawa} and \eqref{eqs:SMYukawa}.
Another remarkable feature is that the tree-level $\mu$ term of $\mu H_u H_d \subset \mu  \mathbf{\bar 6_{H\,\mathrm{I}} }  \mathbf{15_H}$ is impossible in the superpotential~\eqref{eq:SU6W} due to the ${\rm SU}(6)$ gauge symmetry.

\section{The ${\rm SU}(6)$ axion}
\label{sec:SU6axion}
%
%
The physical axion mainly comes from the $ ( \mathbf{1}\,, \mathbf{\bar 3}\,, -\frac{1}{3} )_{\mathbf{H}\, {\rm II}} \supset \frac{v_3}{ \sqrt{2} } e^{i a_3/v_3}$ and the $( \mathbf{1}\,, \mathbf{ 6}\,, +\frac{2}{3} )_{ \mathbf{H} } \supset \frac{v_6}{ \sqrt{2} } e^{i a_6/v_6}$.
It can be obtained from the orthogonality between the ${\rm U}(1)_{\rm PQ}$ current and the ${\rm U}(1)_N$ current, and we arrive at
\beqn
&&- \frac{1}{3} q_3 (v_3)^2 + \frac{2}{3} q_6 (v_6)^2  =0\,.
\eeqn
The physical PQ charges of $(q_3\,, q_6)$ for the $(a_3\,, a_6)$ fields are linear combinations of the ${\rm U}(1)_{\rm PQ}$ charges and the ${\rm U}(1)_N$ charges~\cite{DiLuzio:2020qio}
\beqn\label{eq:qcharge}
q&\equiv& c_1\,{\rm PQ} + c_2\, N\,.
\eeqn
The coefficient of $c_1$ is determined by matching~\cite{tHooft:1979rat} the global anomaly factors of $\gU(1)_{\rm PQ} [\gSU(6)]^2$ and $\gU(1)_{\rm PQ} [\gSU(3)_c]^2$ 
\beqs
\beqn
N_{\gSU(6) }&=& -5 \,, \label{eq:SU6_PQanom} \\
N_{\gSU(3)_c }&=& \sum_{\mR_f \in \gSU(3)_c }  {\rm PQ}_f T(\mR_f)   = - 5 c_1\,, \label{eq:SU3_PQanom}
\eeqn
\eeqs
which leads to $c_1=1$.
By denoting the overall size of the Higgs VEVs and their ratio as
\beqn
&& v_{331}^2 = ( q_3 v_3 )^2 +  ( q_6 v_6 )^2 \,,\quad \tan\phi \equiv \frac{ v_3}{ 2v_6 }\,,
\eeqn
we find the physical PQ charges and axion decay constant of
\beqs
\beqn
&& q_3 = - 3 \cos^2\phi\,,\quad q_6 = -6 \sin^2\phi \,,\\
&& f_a= \frac{ v_{331} }{ 2 | N_{\gSU(3)_c } | } = \frac{ v_{331} }{ 10 } = \frac{ 3 v_3 v_6}{ 5 \sqrt{ (v_3)^2 + 4 ( v_6)^2 } }\,. \label{eq:fa}
\eeqn
\eeqs
The physical axion becomes $a_{\rm phys}=\cos\phi \, a_3 + \sin \phi \, a_6 $.
In addition, the electromagnetic (EM) anomaly factor is
\beqn\label{eq:EM_PQanom}
E&=& \sum_f {\rm PQ}_f {\rm dim} ( \mC_f ) \Tr q_f^2 = -\frac{40}{3} \,,
\eeqn
with $\mC_f$ being the the fermion representations under the $\gSU(3)_c$, and $({\rm PQ}_f \,,q_f)$ being the PQ and EM charges of fermions.

\section{The high-quality axion}
\label{sec:quality}
%
%
The leading PQ-breaking (with $\Delta{\rm PQ}=-12$) operator that is invariant under the ${\rm SU}(6)$, ${\rm SU}(2)_F$, and the discrete $\mathbb{Z}_{4\mR}$ symmetries in Table~\ref{tab:SU6min_PQ} reads
\beqn\label{eq:PQV_Op}
\mO_{ \bcancel{\rm PQ}}^{d=6}&=& \Big( \epsilon^{\rho\delta } \mathbf{\bar 6_{H\,\rho}} \mathbf{\bar 6_{H\,\delta }} \mathbf{15_H} \Big)^2 \supset \Big[  \epsilon^{\rho\delta} \epsilon_{IJK} ( \mathbf{1}\,,  \mathbf{\bar 3}\,, -\frac{1}{3} )_\rho^{I} \non
&& ( \mathbf{1}\,,  \mathbf{\bar 3}\,, -\frac{1}{3})_\delta^{J} ( \mathbf{1}\,,  \mathbf{\bar 3}\,, +\frac{2}{3}  )^{K}   \Big]^2\,,
\eeqn
with $(I\,,J\,,K)$ being the ${\rm SU}(3)_L$ indices.
In order not to reintroduce further PQ-breaking operators, it is reasonable to expect the SUSY-breaking scale to be lower than $v_{331}$.
The total axion effective potential and the induced effective $\bar \theta$ due to the Eq.~\eqref{eq:PQV_Op} are
\beqn\label{eq:AxionPotential}
&& V= V_{\rm QCD} + V_{ \bcancel{\rm PQ}} \,, \quad V_{\rm QCD}=  \Lambda_{\rm QCD}^4 \Big[ 1-  \cos( \frac{ a_{\rm phys} }{f_a} ) \Big] \,, \non
&& V_{ \bcancel{\rm PQ}} = k \frac{ \mO_{ \bcancel{\rm PQ}}^{d=6} }{ M_{\rm pl}^2}+ H.c. \non
&\approx& \frac{ |k| (v_u v_d v_3 )^2 }{ 4 M_{\rm pl}^2} \cos\Big(  \frac{ a_{\rm phys} }{ 6f_a} + \delta \Big) \,,  
\eeqn
with $\delta ={\rm Arg}(k)$ and $M_{\rm pl} = 1.22 \times 10^{19}\,\GeV$.
According to the PQ quality requirement~\cite{Barr:1992qq,Kamionkowski:1992mf,Holman:1992us}, the contribution from the $V_{ \bcancel{\rm PQ}}$ to the energy density should be $10^{-10}$ times less than that of the QCD axion potential, and we find
\beqn\label{eq:PQfa_lim}
f_a & \lesssim &  \frac{3.7\times 10^7\, \cos\phi }{|k \sin (\delta) |^{1/2}} ( \tan\beta + \frac{1}{ \tan\beta} ) \, \GeV \,.
\eeqn
%

\begin{figure}
\begin{minipage}[t]{\hsize}
\includegraphics[width=7.2cm]{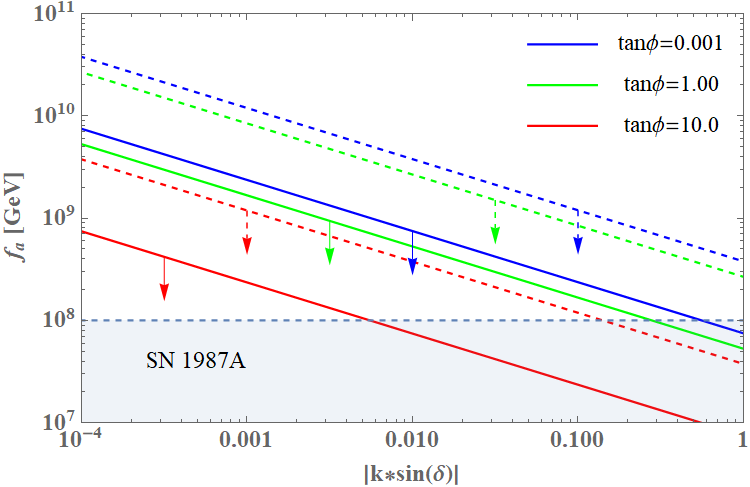}
  \end{minipage}
 \caption{
 The upper limit to the $f_a$ from the dimension-six operator~\eqref{eq:PQV_Op}, with $\tan\beta=1.0$ (solid lines) and $\tan\beta=10.0$ (dashed lines).
 The shaded region represents the lower bound to the $f_a$ from the supernovae 1987A neutrino burst~\cite{Chang:2018rso}, and arrows indicate that each line represents the upper bound to $f_a$.
}
\label{fig:fa} 
\end{figure}

In Fig.~\ref{fig:fa}, we display upper limits to $f_a$.
Within the reasonable parameter choices of $(\tan\beta\,,\tan\phi\,, | k \sin(\delta) |)$, we consider the high-quality axion window of
\beqn\label{eq:highq_fa}
&& 10^{8}\,{\rm GeV} \lesssim f_a \lesssim 10^{10}\,{\rm GeV}  \,,\non
&\Rightarrow& 10^{9}\,{\rm GeV} \lesssim v_{331} \lesssim 10^{11}\,{\rm GeV}  \,,
\eeqn
by using the relation in Eq.~\eqref{eq:fa}.
The corresponding axion masses~\cite{Weinberg:1977ma,Georgi:1986df} are
\beqn\label{eq:ma}
m_a&=& 5.70 \Big( \frac{10^{12}\,\GeV }{f_a}  \Big) \,\mu{\rm eV } \sim (10^{-4}\,, 10^{-2} ) {\rm eV} \,.
\eeqn
This is distinguishable from the axion masses of $m_a\sim \mO(10^{-9})\,{\rm eV}$ in the ${\rm SU}(5)$ GUT~\cite{Boucenna:2017fna,DiLuzio:2018gqe,FileviezPerez:2019fku,FileviezPerez:2019ssf}.

\begin{figure}
\begin{minipage}[t]{\hsize}
\includegraphics[width=8cm]{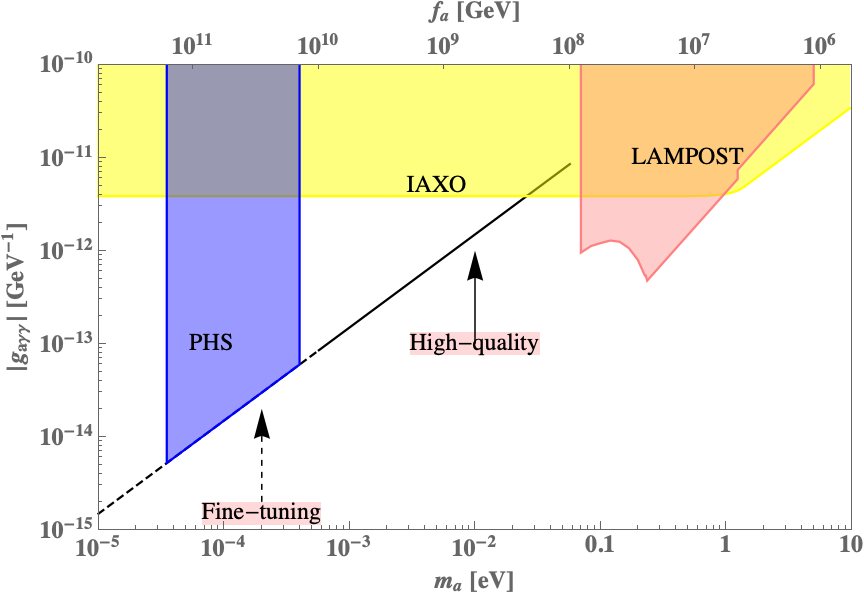}
  \end{minipage}
 \caption{
 The $(m_a\,, |g_{a\gamma\gamma}| )$ predictions for the high-quality ${\rm SU}(6)$ axion (solid line), together with the future projected search limits from the IAXO~\cite{Armengaud:2014gea,Armengaud:2019uso,Ge:2020zww}, LAMPOST~\cite{Baryakhtar:2018doz}, and Plasma halo scope (PHS)~\cite{Lawson:2019brd}.
}
\label{fig:window} 
\end{figure}

The effective axion-photon couplings are parametrized by
\beqn\label{eq:effcouplings}
g_{a\gamma\gamma}&=&  C_{a\gamma\gamma} \Big( \frac{1.14\times  10^{-3}\,\GeV }{f_a}  \Big) \, \GeV^{-1}\,, \non
C_{a\gamma\gamma}&=& \frac{E}{ N_{\gSU(3)_c} } -1.92 \,. \label{eq:aphoton_coup}
\eeqn
By using the QCD and EM anomaly factors in Eqs.~\eqref{eq:SU3_PQanom} and \eqref{eq:EM_PQanom}, we find that $C_{a\gamma\gamma}= 0.75$.
Thus, the axion-photon couplings from the ${\rm SU}(6)$ GUT confirm the universal predictions as those in the ${\rm SU}(5)$ and ${\rm SO}(10)$ GUTs~\cite{FileviezPerez:2019fku,FileviezPerez:2019ssf,DiLuzio:2020qio,DiLuzio:2020wdo}.
In Fig.~\ref{fig:window}, we present the benchmark models for the ${\rm SU}(6)$ axion in the $(m_a\,, |g_{a\gamma\gamma}| )$ plane.
The high-quality axion (solid line) with mass range in Eq.~\eqref{eq:ma} can be probed in the future IAXO~\cite{Armengaud:2014gea,Armengaud:2019uso,Ge:2020zww} experiment.

\section{The axion domain wall}
\label{sec:DW}
%
%
The $V_{ \bcancel{\rm PQ}}$ term in Eq.~\eqref{eq:AxionPotential} plays a role as a biased term~\cite{Vilenkin:1981zs,Gelmini:1988sf,Larsson:1996sp,Kawasaki:2004yh,Kawasaki:2004qu,Saikawa:2017hiv} to avoid the axion domain wall formation, which can be possible due to the periodicity of the effective potential term of $V_{\rm QCD}$.
One thus requires domain walls to decay before the domination epoch~\cite{Sikivie:1982qv}
\beqn\label{eq:DWlim}
&&t_{\rm dec} < t_{\rm form}\,, t_{\rm dec}\sim 10^{- 66}\,{\rm sec}\Big(  \frac{ M_{\rm pl} v_{331}  }{ v_u v_d} \Big)^2  \Big( \frac{ 10^{13}\,\GeV  }{  v_{331} } \Big)^{3} \,, \non
&& t_{\rm form} \sim 10^2\,{\rm sec} \Big( \frac{  10^{13}\,\GeV }{ v_{331}  } \Big) \,, 
\eeqn
where we use the axion domain wall tension of $\sigma_{\rm DW} \approx 9 m_a f_a^2$~\cite{Huang:1985tt,Hiramatsu:2012sc} and Eq.~\eqref{eq:fa}.
With the high-quality range in Eq.~\eqref{eq:highq_fa}, we find that $t_{\rm dec}\sim \mO(10^{-8}) - \mO(10^{-6}) \,{\rm sec}$ and $t_{\rm form} \sim \mO(10^4) - \mO(10^6)\,{\rm sec}$.
Indeed, the axion domain wall cannot be formed in the early Universe.

\section{The gauge coupling unification}
\label{sec:unification}
%
%
We briefly present the gauge coupling evolutions in terms of the one-loop renormalization group equations (RGEs).
The gauge couplings are $( \alpha_{3c} \,, \alpha_{3L} \,, \alpha_N )$ for the $\mG_{331}$ symmetry, and $( \alpha_{3c} \,, \alpha_{2L} \,, \alpha_Y )$ for the $\mG_{\rm SM}$ symmetry.
The ${\rm U}(1)_N$ coupling should be normalized by $\alpha_1 = \frac{4}{3} \alpha_N$ for the unification.

The most general one-loop RGE solution for the gauge coupling $\alpha_i$ of the gauge symmetry $\mG_i$ is 
\beqn
\alpha_i^{-1}( \mu_2) &=& \alpha_i^{-1}( \mu_1 ) - \frac{ b_i^{(1)} }{ 2\pi} \log \Big( \frac{ \mu_2}{ \mu_1} \Big) \,, 
\eeqn
with the one-loop $\beta$ coefficients in the SUSY extension being
\beqn
b_i^{(1)}&=& - 3 C_2( \mG_i ) + \sum_\chi T(\mR_\chi^i )\,.
\eeqn
Explicitly, the one-loop $\beta$ coefficients for the SUSY $\gSU(6)$ can be obtained by the spectrum in Table~\ref{tab:SU6_ferm} and the branching rules in Eq.~\eqref{eqs:SU6to331_Higgs} as follows
\beqn
m_Z\leq \mu \leq v_{331} ~&:&~(b_{ \gSU(3)_c }^{(1)} \,, b_{\gSU(2)_L }^{(1)} \,, b_{\gU(1)_Y}^{(1)} ) \non
& =&  ( -3 \,, 1 \,, 11) \,, \non
v_{331} \leq \mu \leq \Lambda_{\rm GUT}~&:&~ (b_{ \gSU(3)_c }^{(1)} \,, b_{\gSU(3)_L }^{(1)} \,, b_{\gU(1)_1 }^{(1)} ) \non
&=&  ( 0 \,, \frac{13}{2} \,, \frac{29}{2} ) \,.
\eeqn
To evaluate the RGEs, the tree-level matching conditions at the $v_{331}$ scale are
\beqn\label{eq:matching}
&& \alpha_{3L}^{-1} (v_{331} ) = \alpha_{2L}^{-1} (v_{331} ) \,,\non
&& \alpha_{1}^{-1}(v_{331} )  = -\frac{1}{4} \alpha_{2L}^{-1} (v_{331} )  + \frac{3}{4} \alpha_Y^{-1} (v_{331} ) \,.
\eeqn

\begin{figure}
\begin{minipage}[t]{\hsize}
\includegraphics[width=7.5cm]{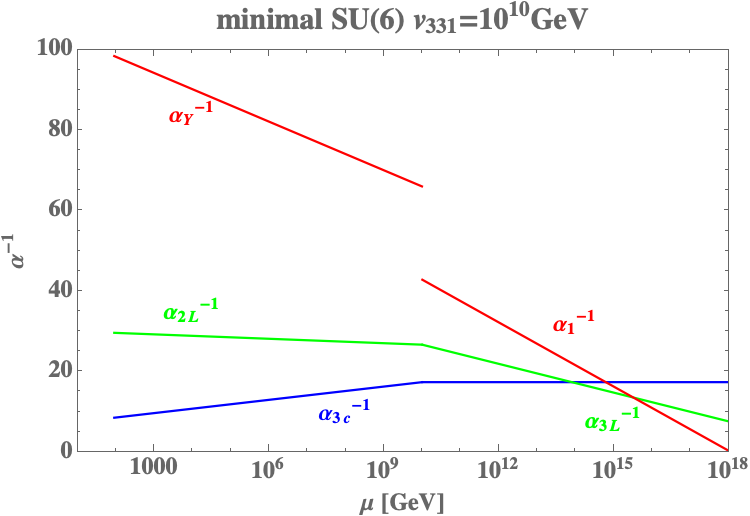}
\end{minipage}
\caption{ The one-loop gauge coupling runnings for the SUSY $\gSU(6)$ model, with $v_{331}= 10^{10}\,\GeV$. }
\label{fig:unification} 
\end{figure}

The one-loop results are displayed in Fig.~\ref{fig:unification}, with an intermediate scale of $v_{331}=10^{10}\,\GeV$ and the latest electroweak precision measurements of $(\alpha_{3c}\,, \alpha_{\rm em}\,, \sin^2\theta_W)$ at the $Z$ pole~\cite{Tanabashi:2018oca} as the inputs.
A unification is indicated with $\Lambda_{\rm GUT}\sim 10^{16}\,\GeV$ for the SUSY ${\rm SU}(6)$.
Notice that the intermediate $\mG_{331}$-breaking scale usually requires the two-loop RGE analysis as well as the one-loop matching conditions with mass threshold effects~\cite{Weinberg:1980wa,Hall:1980kf}.
Recent studies on the ${\rm SO}(10)$ and $E_6$ reveal the strong correlation to the proton lifetime predictions with these effects~\cite{Chakrabortty:2019fov,Meloni:2019jcf,Dash:2020jlc,Ohlsson:2020rjc,King:2021gmj}.
Thus, we defer the study of the proton lifetime predictions with the current constraint of $\tau_p\gtrsim 2.4 \times 10^{34} \,{\rm yrs}$ from the Super-Kamionkande~\cite{Miura:2016krn,Takenaka:2020vqy} to future work.

\section{Conclusions}
\label{sec:conclusion}
%
%
We have put forth a SUSY ${\rm SU}(6)$ model for the strong {\it CP} problem, by utilizing the emergent global DRS symmetry in Eq.~\eqref{eq:DRS_global}.
Historically, the emergent global symmetry was first mentioned in the study of strongly coupled theories.
Its emergence and breaking are independent of the dynamical aspects of the gauge theories.
A high-quality axion with its decay constant is found to be constrained from the PQ-quality requirement to the dimension-six PQ-breaking operator, and is most likely to be directly probed in the future IAXO searches.
The type-I seesaw mechanism and a successful leptogenesis is also achievable at the corresponding PQ symmetry-breaking scale.
Our work manifests that three seemingly unrelated issues of the strong {\it CP} problem, the neutrino mass origin, and the EW Higgs sector are coherently unified in the ${\rm SU}(6)$ framework.
Though the current study of the ${\rm SU}(6)$ requires additional ingredients of discrete symmetries for suitable PQ-breaking operators, it is natural to extend ${\rm SU}(6)$ to higher unified groups, which enjoy the emergent global DRS symmetry in general.
Historically, the {\it nonminimal} GUTs with gauge symmetries of ${\rm SU}(N\geq 7)$ were considered to unify three generations of SM fermions~\cite{Georgi:1979md,Frampton:1979cw,Frampton:1979tj,Frampton:1979fd}.
It is therefore appealing to look for realistic GUTs that unify both the PQ quality problem and the flavor puzzle.

\section*{Acknowledgements}

We would like to thank Lorenzo Calibbi, Gongjun Choi, Luca Di Luzio, Georgi Dvali, Weiqi Fan, Tianjun Li, Zhaofeng Kang, Junle Pei, Yuan Sun, Wenbin Yan, Wen Yin, and Ye-Ling Zhou for very useful discussions and communication during the preparation of this paper. 
N.C. would like to thank Tibet University for hospitality when preparing this work.
We dedicate this work to Dr. Cen Zhang.
This work is partially supported by the National Natural Science Foundation of China (under Grants No. 12035008 and No. 11575176).

\appendix
\section{No high-quality axion in the non-SUSY $\gSU(6)$ model}
\label{sec:proof}

We prove that the non-SUSY $\gSU(6)$ model cannot lead to a high-quality axion with any discrete gauge symmetry of $\mathbb{Z}_k$.
To see this, we assign the most general $\mathbb{Z}_{k}$ charges for fermions and Higgs fields in Table~\ref{tab:SU6_fields}, which are subject to the gauge anomaly cancellation of
\beqn\label{eq:Zk_anomaly}
\mathbb{Z}_{k} [\gSU(6)]^2&=& \sum_i 2 T( \mR_i ) q_i = 2 q_{\bar 6} + 4 q_{15} = 0 ~ {\rm mod}~ k \non
& \Rightarrow & 2 q_{\bar 6} + 4 q_{15} = k\cdot l \,, l \in \mathbb{Z}\,.
\eeqn
In addition, at least two of the following terms are necessary to reproduce the SM-like Higgs boson Yukawa couplings that are consistent with the current LHC results
\beqn\label{eq:miniSU6_Yuk}
\mL_Y&\supset&  \mathbf{15_F}  \mathbf{\bar 6_F^\rho } \mathbf{\bar 6_{H\, \rho} } + \mathbf{ 15_F} \mathbf{ 15_F} \mathbf{ 15_H} + H.c.\,.
\eeqn
The neutrality of the $\mathbb{Z}_k$ charges in Eq.~\eqref{eq:miniSU6_Yuk} leads to 
\beqs
\beqn
&& q_{\bar 6} + q_{15} - q_{6_H} = k\cdot m \,,\label{eq:Zk_neutral1}\\
&&2 q_{15} + q_{15_H} = k\cdot  n \,. \label{eq:Zk_neutral2}
\eeqn
\eeqs
Apparently, the leading PQ-breaking and gauge-invariant operator is $\mO_{ \bcancel{\rm PQ}}^{d=3}=  \epsilon^{\rho\delta} \mathbf{\bar 6_{H\,\rho}} \mathbf{\bar 6_{H\,\delta}}  \mathbf{15_H}$, and it is neutral in $\mathbb{Z}_k$ charge since
\beqn
&& -2 q_{ 6_H } + q_{15_H} = k \cdot ( 2m +n -l ) =0 ~ {\rm mod}~ k
\eeqn
according to Eqs.~\eqref{eq:Zk_anomaly}, \eqref{eq:Zk_neutral1}, and~\eqref{eq:Zk_neutral2}.
The PQ-quality condition from the $\mO_{ \bcancel{\rm PQ}}^{d=3}$ clearly leads to an unrealistic constraint of
\beqn
&&  v_{\rm EW}^2 v_{331} \lesssim 10^{-10} \frac{ \Lambda_{\rm QCD}^4 }{ M_{\rm pl}} \Rightarrow v_{331} \lesssim 10^{-37}\,{\rm GeV} \,.
\eeqn
Thus, a high-quality axion is impossible in the non-SUSY $\gSU(6)$ model.

\begin{table}[htp]
\begin{center}
\begin{tabular}{c|cc|cccc}
\hline
 &  $ \mathbf{ \bar 6_F^\rho} $ & $\mathbf{15_F}$ & $ \mathbf{\bar 6_{H\,\rho}} $ & $\mathbf{15_H}$ & $\mathbf{21_H}$   & $ \mathbf{35_H} $   \\
\hline
$\gSU(2)_F$  &  $\Box$  & $1$  &  $\overline\Box$  & $1$  & $1$  & $1$   \\
$\gU(1)_{\rm PQ}$  & $1$ & $1$ & $-2$ & $- 2$ & $-2 $ &  $0$    \\
$\mathbb{Z}_{k}$  &  $q_{\bar 6}$  & $q_{15}$  & $q_{6_H}$ & $q_{15_H}$  & $q_{21_H}$  & $q_{35_H}$   \\
\hline
\end{tabular}
\end{center}
\caption{The fermions and Higgs fields with their $\gSU(2)_F$ representations, and the charges under the $\gU(1)_{\rm PQ}$ and $\mathbb{Z}_{k}$ symmetries, in the non-SUSY $\gSU(6)$ model.}
\label{tab:SU6_fields}
\end{table}%

\section{The $\gSU(6)$ gauge symmetry breaking and spectrum}
\label{sec:SU6spectrum}

In this section, we list the SUSY $\gSU(6)$ spectrum and the Yukawa couplings following the breaking pattern of $\gSU(6) \to \mG_{331}\to \mG_{\rm SM}$.
The $\gU(1)_N$ charge for the $\gSU(6)$ fundamental representation at the first-stage symmetry breaking is defined as follows
\beqn\label{eq:Ncharge}
N&\equiv& \frac{1}{3}{\rm diag} (-1\,,-1\,,-1\,,+1\,,+1\,,+1) \non
&=& -\frac{ \sqrt{6} }{3} T_{ \gSU(6) }^{15 } -  \frac{ \sqrt{10} }{5 } T_{ \gSU(6) }^{24}- \frac{ 2}{ \sqrt{15} }T_{ \gSU(6) }^{35}\,.
\eeqn
Afterwards, the $\gU(1)_Y$ charge for the $\gSU(3)_L$ fundamental representation is given by
\beqn\label{eq:U1Y}
Y&\equiv&  {\rm diag} ( \frac{1}{3} + 2N \,,  \frac{1}{3} + 2N \,, - \frac{2}{3} + 2N  ) \non
& =& \frac{2}{ \sqrt{3}} T_{ \gSU(3) }^8 + 2 N\cdot \mathbb{I}_3 \,,
\eeqn
and the electric charges are quantized by
\beqn
Q&\equiv& T_{ \gSU(2) }^3 + \frac{ Y}{2} \mathbb{I}_2 \,.
\eeqn
$T_{ \gSU(6) }^{15\,,24\,, 35} $, $T_{ \gSU(3) }^8$, and $T_{ \gSU(2) }^3$ are Cartan generators of $\gSU(6)$, $\gSU(3)_L$, and $\gSU(2)_L$, respectively.
The fermion spectrum from the superfields of $\mathbf{\bar 6_F^\rho} \oplus \mathbf{15_F} $ are tabulated in Table~\ref{tab:SU6_ferm}, with their representations under the $\mG_{331}$ and $\mG_{\rm SM}$.
The first-stage branching rules of superfields containing Higgs are
\beqs\label{eqs:SU6to331_Higgs}
\beqn
\mathbf{\bar 6_{H\,\rho} }  &=& (\mathbf{\bar 3}\,, \mathbf{1}\,, + \frac{1}{3})_{ \mathbf{H}\,\rho } \oplus ( \mathbf{1}\,, \mathbf{\bar 3}\,,-\frac{1}{3} )_{ \mathbf{H}\,\rho } \,,\\
\mathbf{15_H} &=& (\mathbf{\bar 3}\,, \mathbf{1}\,, - \frac{2}{3})_{ \mathbf{H} } \oplus ( \mathbf{1} \,, \mathbf{\bar 3} \,, + \frac{2}{3} )_{\mathbf{H}} \non
&\oplus& ( \mathbf{3} \,, \mathbf{ 3}\,,0 )_{\mathbf{H}} \,,\\
 \mathbf{21_H} &=& (\mathbf{6}\,, \mathbf{1}\,, - \frac{2}{3} )_{ \mathbf{H} }  \oplus  (\mathbf{1}\,, \mathbf{6}\,, +\frac{2}{3} )_{ \mathbf{H} }  \non
 &\oplus& (\mathbf{3}\,, \mathbf{3}\,, 0)_{ \mathbf{H}}^\prime \,,\\
 \mathbf{\overline{21}_H} &=& (\mathbf{\bar 6}\,, \mathbf{1}\,, +\frac{2}{3} )_{ \mathbf{H} }  \oplus  (\mathbf{1}\,, \mathbf{\bar 6}\,, -\frac{2}{3} )_{ \mathbf{H} } \non  
 &\oplus& (\mathbf{\bar 3}\,, \mathbf{\bar 3}\,, 0)_{ \mathbf{H}}^\prime \,,\\
\mathbf{35_H}  &=& ( \mathbf{1}  \,, \mathbf{1} \,,0 )_{ \mathbf{H} }  \oplus ( \mathbf{8}  \,, \mathbf{1} \,,0 )_{\mathbf{H}} \oplus ( \mathbf{1}  \,, \mathbf{8} \,,0 )_{\mathbf{H}} \non
& \oplus& ( \mathbf{3}  \,, \mathbf{\bar 3} \,,- \frac{2}{3} )_{\mathbf{H} }\oplus ( \mathbf{ \bar 3}  \,, \mathbf{3} \,,+ \frac{2}{3} )_{ \mathbf{H} }  \,.
\eeqn
\eeqs
The scalar components of $( \mathbf{1}\,, \mathbf{\bar 3}\,, -\frac{1}{3} )_{ \mathbf{H}\,{\rm II}}  \subset \mathbf{\bar 6_{H\,{\rm II}}  } $ and the $(\mathbf{1}\,, \mathbf{6}\,, +\frac{2}{3} )_{ \mathbf{H} }   \subset  \mathbf{21_H}$ will develop VEVs of $\sim v_{331}$ for the symmetry breaking of $\mG_{331} \to \mG_{\rm SM}$.
We can determine the global $\gU(1)_{\rm PQ} [\gSU(6)]^2$ anomaly factor from the fermions in Table~\ref{tab:SU6_ferm} and Eq.~\eqref{eqs:SU6to331_Higgs} as follows
\beqn\label{eq:SU6_PQanom_derive}
N_{\gSU(6) }&=& \Big[  \sum_\rho T(\mathbf{\bar 6_F^\rho }) {\rm PQ} ( \mathbf{\bar 6_F^\rho }) +  T(\mathbf{15_F}) {\rm PQ} ( \mathbf{15_F} ) \Big] \times n_g  \non
&+& \sum_\rho T(\mathbf{ \bar 6_{H\,\rho}}) {\rm PQ} ( \mathbf{\bar 6_{H\,\rho} } ) + T(\mathbf{ 15_H}) {\rm PQ} ( \mathbf{ 15_H } ) \non
&+& T(\mathbf{ 21_H}) {\rm PQ} ( \mathbf{ 21_H } )  =  -5 \,,
\eeqn
with $n_g=3$.
The related trace invariants are $T( \mathbf{6 } ) = T( \mathbf{\bar 6 } ) =\frac{1}{2}$, $T(\mathbf{ 15 })=2$, and $T(\mathbf{ 21 })=4$.

After the second-stage symmetry breaking, we find the following mass terms from the Yukawa couplings 
\beqs\label{eqs:331Yukawa}
\beqn\label{eq:Dmass}
&& Y_D \mathbf{15_F} \mathbf{\bar 6_F^{\rm II }}  \mathbf{\bar 6_{H\,{\rm II}} }  + H.c. \supset  \non
&& Y_D \Big[  ( \mathbf{3}\,, \mathbf{ 3}\,, 0 )_{ \mathbf{F}}  \otimes ( \mathbf{\bar 3}\,, \mathbf{ 1}\,, +\frac{1}{3} )_{ \mathbf{F}}^{ \mathrm{II}} \non
& \oplus&  (  \mathbf{ 1}\,, \mathbf{\bar 3}\,, +\frac{2}{3} )_{ \mathbf{F}} \otimes (  \mathbf{ 1}\,, \mathbf{\bar 3}\,, -\frac{1}{3} )_{ \mathbf{F}}^{ \mathrm{II}}  \Big]\otimes ( \mathbf{1}\,, \mathbf{ \bar 3}\,,  -\frac{1}{3} )_{ \mathbf{H}\,{\rm II}}+ H.c. \non
&\Rightarrow& m_D = m_{e^\prime} = m_{\nu^\prime } = Y_D\, v_{3}  \,, \\
&& Y_N \mathbf{\bar 6_F^{[ \mathrm{I } }}  \mathbf{\bar 6_F^{\mathrm{II } ] }}  \mathbf{21_H } + H.c. \supset  \non
&& Y_N ( \mathbf{1}\,, \mathbf{ \bar 3}\,,  -\frac{1}{3} )_{ \mathbf{F}}^{ \mathrm{[ I}} \otimes ( \mathbf{1}\,, \mathbf{ \bar 3}\,,  -\frac{1}{3} )_{ \mathbf{F}}^{ \mathrm{II ]}} \otimes  ( \mathbf{1}\,, \mathbf{  6}\,,  +\frac{2}{3} )_{ \mathbf{H}}  + H.c.  \non
&\Rightarrow& m_{N\,, N^\prime} =  Y_N v_{6} \,.
\eeqn
\eeqs
The global $\gU(1)_{\rm PQ} [\gSU(3)_c]^2$ anomaly factor is determined as follows
\beqn\label{eq:SU3_PQanom_derive}
N_{\gSU(3)_c } &=& \small{ \Big[ \underbrace{ \sum_\rho T( \mathbf{\bar 3_F^\rho } ) ( c_1 + \frac{1}{3} c_2) }_{ \mathbf{\bar 6_F^\rho}} } \non
& +& \small{ \underbrace{ T( \mathbf{\bar 3_F} ) ( c_1 - \frac{2}{3} c_2 ) + T( \mathbf{ 3_F} )  3 c_1 }_{ \mathbf{15_F} } \Big]  \times n_g } \non
&+&\small{  \underbrace{ \sum_\rho T( \mathbf{\bar 3_{H\,\rho} } ) (- 2c_1 + \frac{1}{3} c_2 ) }_{  \mathbf{\bar 6_{H\,\rho}}} }\non
&+& \small{ \underbrace{T( \mathbf{\bar 3_H} ) (- 2c_1 - \frac{2}{3} c_2 ) + T( \mathbf{3_H} ) (-6c_1) }_{ \mathbf{15_H}}  } \non
&+& \small{  \underbrace{ T( \mathbf{ 6_H} ) ( - 2c_1 + \frac{2}{3} c_2 ) + T( \mathbf{3_H} ) (-6c_1)}_{ \mathbf{21_H}} }\non
&+& \small{  \underbrace{ T( \mathbf{\bar 6_H } ) (  - \frac{2}{3} c_2 ) }_{ \mathbf{ \overline{21}_H } } = - 5c_1 \,,}
\eeqn
according to the physical PQ charges of $q\equiv c_1 {\rm PQ} + c_2 N$.

\begin{table}[htp]
\begin{center}
\begin{tabular}{c|c|c}
\hline \hline
   $\gSU(6)$   &  $\mG_{331}$  & $\mG_{\rm SM}$  \\
\hline \hline
  $\mathbf{\bar 6_F^{\mathrm{I} }} $ & $(\mathbf{\bar 3} \,, \mathbf{1} \,, + \frac{1}{3})_{ \mathbf{F} }^{\mathrm{I} }$    &  $(\mathbf{\bar 3} \,, \mathbf{1} \,, + \frac{2}{3})_{ \mathbf{F} }^{\mathrm{I} }~:~ \underline{ d_R^c } $ \\   
    & $(\mathbf{1} \,, \mathbf{\bar 3} \,,  -\frac{1}{3})_{ \mathbf{F} }^{\mathrm{I} }$  &  $( \mathbf{1} \,, \mathbf{ 2}  \,, -1 )_{ \mathbf{F} }^{\mathrm{I} }~:~ \underline{ ( e_L \,, -\nu_L)} $\\ 
    &  & $( \mathbf{1} \,, \mathbf{ 1}  \,, 0 )_{ \mathbf{F} }^{\mathrm{I} }~:~ \dotuline{ N }$  \\ \hline
 $\mathbf{\bar 6_F^{\mathrm{II} } }$  & $(\mathbf{\bar 3} \,, \mathbf{1} \,, + \frac{1}{3})_{ \mathbf{F} }^{\mathrm{II} }$   &  $(\mathbf{\bar 3} \,, \mathbf{1} \,, + \frac{2}{3})_{ \mathbf{F} }^{\mathrm{II} }~:~ \dashuline{D_R^c}$ \\
    &  $(\mathbf{1} \,, \mathbf{\bar 3} \,,  -\frac{1}{3})_{ \mathbf{F} }^{\mathrm{II}}$ &  $( \mathbf{1} \,, \mathbf{ 2}  \,, -1 )_{ \mathbf{F} }^{\mathrm{II} } ~:~ \dotuline{ ( e_L^\prime \,, - \nu_L^\prime )} $\\ 
   &   &  $( \mathbf{1} \,, \mathbf{ 1}  \,, 0 )_{ \mathbf{F} }^{\mathrm{II} }~:~ \dotuline{ N^\prime }$  \\ \hline
    $\mathbf{15_F}$ & $(\mathbf{\bar 3} \,, \mathbf{1} \,, - \frac{2}{3})_{ \mathbf{F}}$  & $(\mathbf{\bar 3} \,, \mathbf{1} \,, - \frac{4}{3})_{ \mathbf{F} }~:~ \underline{u_R^c}$ \\ 
    & $( \mathbf{1}\,, \mathbf{\bar 3} \,, + \frac{2}{3})_{ \mathbf{F}}$  & $( \mathbf{1} \,, \mathbf{2}  \,, +1)_{ \mathbf{F} }~:~ \dotuline{ ( \nu_R^{\prime\, c} \,, e_R^{\prime\, c} )}$ \\ 
  &    & $( \mathbf{1} \,, \mathbf{1}  \,, +2)_{ \mathbf{F} }~:~\underline{e_R^c} $ \\  
  & $(\mathbf{ 3} \,, \mathbf{3} \,, 0)_{ \mathbf{F} }$  &    $(\mathbf{ 3} \,, \mathbf{2} \,, +\frac{1}{3})_{ \mathbf{F} }~:~ \underline{ (u_L\,, d_L) }$ \\  
  &   &   $(\mathbf{ 3} \,, \mathbf{1} \,, -\frac{2}{3} )_{ \mathbf{F} }~:~ \dashuline{ D_L } $ \\ \hline
\hline
\end{tabular}
\end{center}
\caption{
The $\gSU(6)$ fermion representations under the $\mG_{331}$ and the $\mG_{\rm SM}$.
The SM fermions are marked by solid underlines, the Kim-Shifman-Vainshtein-Zakharov vectorlike quarks are marked by dashed underlines, and other heavy leptonic states are marked by dotted underlines.}
\label{tab:SU6_ferm}
\end{table}%

The scalar components of $( \mathbf{1}\,, \mathbf{2}\,, -1 )_{ \mathbf{H}\,{\rm I}}  \subset ( \mathbf{1}\,, \mathbf{\bar 3}\,,-\frac{1}{3} )_{ \mathbf{H}\,{\rm I}} \subset \mathbf{\bar 6_{H\,{\rm I}}}$ and the $( \mathbf{1}\,, \mathbf{2}\,,+1 )_{ \mathbf{H} } \subset ( \mathbf{1}\,, \mathbf{\bar 3}\,,+\frac{2}{3} )_{ \mathbf{H}  } \subset \mathbf{15_H  }$ will develop VEVs of $\sim v_{\rm EW}$ for the EW symmetry breaking.
The corresponding Yukawa couplings and mass terms are
\beqs\label{eqs:SMYukawa}
\beqn
&& Y_d \mathbf{15_F} \mathbf{\bar 6_F^{\mathrm{I } }}  \mathbf{\bar 6_{H\, {\rm I}} } + H.c. \supset \non
&&  Y_d \Big[ ( \mathbf{ 3}\,, \mathbf{ 2}\,, +\frac{1}{3} )_{\mathbf{F} } \otimes ( \mathbf{\bar 3}\,, \mathbf{ 1}\,, +\frac{2}{3} )_{\mathbf{F}}^{\mathrm{I} } \non
&\oplus& ( \mathbf{ 1}\,, \mathbf{ 2}\,, -1)_{\mathbf{F}}\otimes ( \mathbf{ 1}\,, \mathbf{ 1}\,, +2)_{\mathbf{F}}^{\mathrm{I}} \Big] \otimes ( \mathbf{ 1}\,, \mathbf{ 2}\,, -1)_{\mathbf{H}\,{\rm I}} + H.c. \non
&\Rightarrow& m_{d\,, \ell} = Y_d\, v_{\rm EW} \,,\\
&& Y_u \mathbf{15_F}\mathbf{15_F} \mathbf{15_H} + H.c. \supset \non
&& Y_u ( \mathbf{ 3}\,, \mathbf{ 2}\,, +\frac{1}{3})_{\mathbf{F}} \otimes  ( \mathbf{ \bar 3}\,, \mathbf{1}\,, -\frac{4}{3})_{\mathbf{F}} \otimes  ( \mathbf{ 1}\,, \mathbf{ 2}\,, +1)_{\mathbf{H}} + H.c. \non
&\Rightarrow & m_u  =  (Y_u+ Y_u^T)\, v_{\rm EW }\,,\\
&& Y_\nu \mathbf{\bar 6_F^{ [ {\rm I } }}  \mathbf{\bar 6_F^{ {\rm II} ] }} \mathbf{15_H } + H.c. \supset \non
&& Y_\nu \Big[ ( \mathbf{ 1}\,, \mathbf{ 2}\,, -1)_{\mathbf{F}}^{ [ {\rm I}  } \otimes ( \mathbf{ 1}\,, \mathbf{1}\,, 0)_{\mathbf{F}}^{ {\rm II} ]} \otimes  ( \mathbf{ 1}\,, \mathbf{ 2}\,, +1)_{\mathbf{H}}  \Big] + H.c. \non
&\Rightarrow& m_{\nu N^\prime  } = m_{ \nu^\prime N } = (Y_\nu + Y_\nu^T)\, v_{\rm EW } \,.
\eeqn
\eeqs
The $ (\mathbf{ 1}\,, \mathbf{ 2}\,, -1 )_{\mathbf{H}\,{\rm I}} \subset \mathbf{\bar 6_{H\,{\rm I}}}$ gives masses to down-type quarks and charged leptons, and $( \mathbf{ 1}\,, \mathbf{ 2}\,, +1)_{\mathbf{H}} \subset \mathbf{15_H} $ gives masses to up-type quarks.
This justifies the low-energy effective theory at the EW scale is the type-II 2HDM.
The neutrino masses are realized through the type-I seesaw mechanism
\beqn
&& ( \nu\,, N^\prime ) \cdot \left(  \ba{cc}  0 &  m_{\nu N^\prime } \\   m_{\nu N^\prime }  &  m_{N^\prime} \\  \ea \right) \cdot  \left(  \ba{c} \nu \\ N^\prime   \ea \right) \non
& \Rightarrow& m_\nu =(Y_\nu + Y_\nu^T)^2 Y_N^{-1} \frac{v_{\rm EW}^2 }{ v_6 } \,.
\eeqn
By combining the fermions in Table~\ref{tab:SU6_ferm} and in Eqs.~\eqref{eqs:SU6to331_Higgs}, we can determine the electromagnetic anomaly factor as
\beqn
E&=& \sum_f {\rm PQ}_f {\rm dim} ( {\cal C}_f ) {\rm Tr} q_f^2 \non
&=& \small{ \Big[  \underbrace{ \Big( 3_c (\frac{1}{3})^2  + (-1)^2 \Big) \times 2 }_{ \mathbf{6_F^\rho}} } \non
& +& \small{  \underbrace{ 3_c ( -\frac{2}{3} )^2 + 1 + 1 + 3_c \times \Big(  (\frac{2}{3} )^2 + (-\frac{1}{3} )^2 + (-\frac{1}{3} )^2 \Big) }_{ \mathbf{15_F} }  \Big] \times n_g }\non
&+& \small{ \underbrace{ (-2) \times \Big[  3_c ( \frac{1}{3} )^2 + (-1)^2 \Big] \times 2 }_{ \mathbf{\bar 6_{H\,\rho} }}  } \non
&+& \small{ \underbrace{ (-2) \times \Big[ 3_c (-\frac{2}{3})^2 + 1\times 2 + 3_c (\frac{2}{3} )^2 + 3_c ( - \frac{1}{3} )^2 \times 2  \Big] }_{ \mathbf{15_H} } } \non
&+& \small{ \underbrace{ (-2) \times \Big[ 6_c (-\frac{2}{3})^2 + 2^2 + 1\times 2 + 3_c (\frac{2}{3} )^2  + 3_c ( - \frac{1}{3} )^2 \times 2  \Big] }_{ \mathbf{21_H} } }\non 
&=& -\frac{40}{3} \,.
\eeqn


\begin{thebibliography}{122}%
\makeatletter
\providecommand \@ifxundefined [1]{%
 \@ifx{#1\undefined}
}%
\providecommand \@ifnum [1]{%
 \ifnum #1\expandafter \@firstoftwo
 \else \expandafter \@secondoftwo
 \fi
}%
\providecommand \@ifx [1]{%
 \ifx #1\expandafter \@firstoftwo
 \else \expandafter \@secondoftwo
 \fi
}%
\providecommand \natexlab [1]{#1}%
\providecommand \enquote  [1]{``#1''}%
\providecommand \bibnamefont  [1]{#1}%
\providecommand \bibfnamefont [1]{#1}%
\providecommand \citenamefont [1]{#1}%
\providecommand \href@noop [0]{\@secondoftwo}%
\providecommand \href [0]{\begingroup \@sanitize@url \@href}%
\providecommand \@href[1]{\@@startlink{#1}\@@href}%
\providecommand \@@href[1]{\endgroup#1\@@endlink}%
\providecommand \@sanitize@url [0]{\catcode `\\12\catcode `\$12\catcode
  `\&12\catcode `\#12\catcode `\^12\catcode `\_12\catcode `\%12\relax}%
\providecommand \@@startlink[1]{}%
\providecommand \@@endlink[0]{}%
\providecommand \url  [0]{\begingroup\@sanitize@url \@url }%
\providecommand \@url [1]{\endgroup\@href {#1}{\urlprefix }}%
\providecommand \urlprefix  [0]{URL }%
\providecommand \Eprint [0]{\href }%
\providecommand \doibase [0]{http://dx.doi.org/}%
\providecommand \selectlanguage [0]{\@gobble}%
\providecommand \bibinfo  [0]{\@secondoftwo}%
\providecommand \bibfield  [0]{\@secondoftwo}%
\providecommand \translation [1]{[#1]}%
\providecommand \BibitemOpen [0]{}%
\providecommand \bibitemStop [0]{}%
\providecommand \bibitemNoStop [0]{.\EOS\space}%
\providecommand \EOS [0]{\spacefactor3000\relax}%
\providecommand \BibitemShut  [1]{\csname bibitem#1\endcsname}%
\let\auto@bib@innerbib\@empty
\bibitem [{\citenamefont {Georgi}\ and\ \citenamefont
  {Glashow}(1974)}]{Georgi:1974sy}%
  \BibitemOpen
  \bibfield  {author} {\bibinfo {author} {\bibfnamefont {H.}~\bibnamefont
  {Georgi}}\ and\ \bibinfo {author} {\bibfnamefont {S.~L.}\ \bibnamefont
  {Glashow}},\ }\href {\doibase 10.1103/PhysRevLett.32.438} {\bibfield
  {journal} {\bibinfo  {journal} {Phys. Rev. Lett.}\ }\textbf {\bibinfo
  {volume} {32}},\ \bibinfo {pages} {438} (\bibinfo {year} {1974})}\BibitemShut
  {NoStop}%
\bibitem [{\citenamefont {Fritzsch}\ and\ \citenamefont
  {Minkowski}(1975)}]{Fritzsch:1974nn}%
  \BibitemOpen
  \bibfield  {author} {\bibinfo {author} {\bibfnamefont {H.}~\bibnamefont
  {Fritzsch}}\ and\ \bibinfo {author} {\bibfnamefont {P.}~\bibnamefont
  {Minkowski}},\ }\href {\doibase 10.1016/0003-4916(75)90211-0} {\bibfield
  {journal} {\bibinfo  {journal} {Annals Phys.}\ }\textbf {\bibinfo {volume}
  {93}},\ \bibinfo {pages} {193} (\bibinfo {year} {1975})}\BibitemShut
  {NoStop}%
\bibitem [{\citenamefont {Pendlebury}\ \emph {et~al.}(2015)\citenamefont
  {Pendlebury} \emph {et~al.}}]{Afach:2015sja}%
  \BibitemOpen
  \bibfield  {author} {\bibinfo {author} {\bibfnamefont {J.~M.}\ \bibnamefont
  {Pendlebury}} \emph {et~al.},\ }\href {\doibase 10.1103/PhysRevD.92.092003}
  {\bibfield  {journal} {\bibinfo  {journal} {Phys. Rev. D}\ }\textbf {\bibinfo
  {volume} {92}},\ \bibinfo {pages} {092003} (\bibinfo {year} {2015})},\
  \Eprint {http://arxiv.org/abs/1509.04411} {arXiv:1509.04411 [hep-ex]}
  \BibitemShut {NoStop}%
\bibitem [{\citenamefont {Peccei}\ and\ \citenamefont
  {Quinn}(1977)}]{Peccei:1977hh}%
  \BibitemOpen
  \bibfield  {author} {\bibinfo {author} {\bibfnamefont {R.~D.}\ \bibnamefont
  {Peccei}}\ and\ \bibinfo {author} {\bibfnamefont {H.~R.}\ \bibnamefont
  {Quinn}},\ }\href {\doibase 10.1103/PhysRevLett.38.1440} {\bibfield
  {journal} {\bibinfo  {journal} {Phys. Rev. Lett.}\ }\textbf {\bibinfo
  {volume} {38}},\ \bibinfo {pages} {1440} (\bibinfo {year}
  {1977})}\BibitemShut {NoStop}%
\bibitem [{\citenamefont {Weinberg}(1978)}]{Weinberg:1977ma}%
  \BibitemOpen
  \bibfield  {author} {\bibinfo {author} {\bibfnamefont {S.}~\bibnamefont
  {Weinberg}},\ }\href {\doibase 10.1103/PhysRevLett.40.223} {\bibfield
  {journal} {\bibinfo  {journal} {Phys. Rev. Lett.}\ }\textbf {\bibinfo
  {volume} {40}},\ \bibinfo {pages} {223} (\bibinfo {year} {1978})}\BibitemShut
  {NoStop}%
\bibitem [{\citenamefont {Wilczek}(1978)}]{Wilczek:1977pj}%
  \BibitemOpen
  \bibfield  {author} {\bibinfo {author} {\bibfnamefont {F.}~\bibnamefont
  {Wilczek}},\ }\href {\doibase 10.1103/PhysRevLett.40.279} {\bibfield
  {journal} {\bibinfo  {journal} {Phys. Rev. Lett.}\ }\textbf {\bibinfo
  {volume} {40}},\ \bibinfo {pages} {279} (\bibinfo {year} {1978})}\BibitemShut
  {NoStop}%
\bibitem [{\citenamefont {Kim}(1979)}]{Kim:1979if}%
  \BibitemOpen
  \bibfield  {author} {\bibinfo {author} {\bibfnamefont {J.~E.}\ \bibnamefont
  {Kim}},\ }\href {\doibase 10.1103/PhysRevLett.43.103} {\bibfield  {journal}
  {\bibinfo  {journal} {Phys. Rev. Lett.}\ }\textbf {\bibinfo {volume} {43}},\
  \bibinfo {pages} {103} (\bibinfo {year} {1979})}\BibitemShut {NoStop}%
\bibitem [{\citenamefont {Shifman}\ \emph {et~al.}(1980)\citenamefont
  {Shifman}, \citenamefont {Vainshtein},\ and\ \citenamefont
  {Zakharov}}]{Shifman:1979if}%
  \BibitemOpen
  \bibfield  {author} {\bibinfo {author} {\bibfnamefont {M.~A.}\ \bibnamefont
  {Shifman}}, \bibinfo {author} {\bibfnamefont {A.~I.}\ \bibnamefont
  {Vainshtein}}, \ and\ \bibinfo {author} {\bibfnamefont {V.~I.}\ \bibnamefont
  {Zakharov}},\ }\href {\doibase 10.1016/0550-3213(80)90209-6} {\bibfield
  {journal} {\bibinfo  {journal} {Nucl. Phys. B}\ }\textbf {\bibinfo {volume}
  {166}},\ \bibinfo {pages} {493} (\bibinfo {year} {1980})}\BibitemShut
  {NoStop}%
\bibitem [{\citenamefont {Zhitnitsky}(1980)}]{Zhitnitsky:1980tq}%
  \BibitemOpen
  \bibfield  {author} {\bibinfo {author} {\bibfnamefont {A.~R.}\ \bibnamefont
  {Zhitnitsky}},\ }\href@noop {} {\bibfield  {journal} {\bibinfo  {journal}
  {Sov. J. Nucl. Phys.}\ }\textbf {\bibinfo {volume} {31}},\ \bibinfo {pages}
  {260} (\bibinfo {year} {1980})}\BibitemShut {NoStop}%
\bibitem [{\citenamefont {Dine}\ \emph {et~al.}(1981)\citenamefont {Dine},
  \citenamefont {Fischler},\ and\ \citenamefont {Srednicki}}]{Dine:1981rt}%
  \BibitemOpen
  \bibfield  {author} {\bibinfo {author} {\bibfnamefont {M.}~\bibnamefont
  {Dine}}, \bibinfo {author} {\bibfnamefont {W.}~\bibnamefont {Fischler}}, \
  and\ \bibinfo {author} {\bibfnamefont {M.}~\bibnamefont {Srednicki}},\ }\href
  {\doibase 10.1016/0370-2693(81)90590-6} {\bibfield  {journal} {\bibinfo
  {journal} {Phys. Lett. B}\ }\textbf {\bibinfo {volume} {104}},\ \bibinfo
  {pages} {199} (\bibinfo {year} {1981})}\BibitemShut {NoStop}%
\bibitem [{\citenamefont {Wise}\ \emph {et~al.}(1981)\citenamefont {Wise},
  \citenamefont {Georgi},\ and\ \citenamefont {Glashow}}]{Wise:1981ry}%
  \BibitemOpen
  \bibfield  {author} {\bibinfo {author} {\bibfnamefont {M.~B.}\ \bibnamefont
  {Wise}}, \bibinfo {author} {\bibfnamefont {H.}~\bibnamefont {Georgi}}, \ and\
  \bibinfo {author} {\bibfnamefont {S.~L.}\ \bibnamefont {Glashow}},\ }\href
  {\doibase 10.1103/PhysRevLett.47.402} {\bibfield  {journal} {\bibinfo
  {journal} {Phys. Rev. Lett.}\ }\textbf {\bibinfo {volume} {47}},\ \bibinfo
  {pages} {402} (\bibinfo {year} {1981})}\BibitemShut {NoStop}%
\bibitem [{\citenamefont {Lazarides}(1982)}]{Lazarides:1981kz}%
  \BibitemOpen
  \bibfield  {author} {\bibinfo {author} {\bibfnamefont {G.}~\bibnamefont
  {Lazarides}},\ }\href {\doibase 10.1103/PhysRevD.25.2425} {\bibfield
  {journal} {\bibinfo  {journal} {Phys. Rev. D}\ }\textbf {\bibinfo {volume}
  {25}},\ \bibinfo {pages} {2425} (\bibinfo {year} {1982})}\BibitemShut
  {NoStop}%
\bibitem [{\citenamefont {Georgi}\ \emph {et~al.}(1981)\citenamefont {Georgi},
  \citenamefont {Hall},\ and\ \citenamefont {Wise}}]{Georgi:1981pu}%
  \BibitemOpen
  \bibfield  {author} {\bibinfo {author} {\bibfnamefont {H.~M.}\ \bibnamefont
  {Georgi}}, \bibinfo {author} {\bibfnamefont {L.~J.}\ \bibnamefont {Hall}}, \
  and\ \bibinfo {author} {\bibfnamefont {M.~B.}\ \bibnamefont {Wise}},\ }\href
  {\doibase 10.1016/0550-3213(81)90433-8} {\bibfield  {journal} {\bibinfo
  {journal} {Nucl. Phys. B}\ }\textbf {\bibinfo {volume} {192}},\ \bibinfo
  {pages} {409} (\bibinfo {year} {1981})}\BibitemShut {NoStop}%
\bibitem [{\citenamefont {Rubakov}(1997)}]{Rubakov:1997vp}%
  \BibitemOpen
  \bibfield  {author} {\bibinfo {author} {\bibfnamefont {V.~A.}\ \bibnamefont
  {Rubakov}},\ }\href {\doibase 10.1134/1.567390} {\bibfield  {journal}
  {\bibinfo  {journal} {JETP Lett.}\ }\textbf {\bibinfo {volume} {65}},\
  \bibinfo {pages} {621} (\bibinfo {year} {1997})},\ \Eprint
  {http://arxiv.org/abs/hep-ph/9703409} {arXiv:hep-ph/9703409} \BibitemShut
  {NoStop}%
\bibitem [{\citenamefont {Co}\ \emph {et~al.}(2016)\citenamefont {Co},
  \citenamefont {D'Eramo},\ and\ \citenamefont {Hall}}]{Co:2016xti}%
  \BibitemOpen
  \bibfield  {author} {\bibinfo {author} {\bibfnamefont {R.~T.}\ \bibnamefont
  {Co}}, \bibinfo {author} {\bibfnamefont {F.}~\bibnamefont {D'Eramo}}, \ and\
  \bibinfo {author} {\bibfnamefont {L.~J.}\ \bibnamefont {Hall}},\ }\href
  {\doibase 10.1103/PhysRevD.94.075001} {\bibfield  {journal} {\bibinfo
  {journal} {Phys. Rev. D}\ }\textbf {\bibinfo {volume} {94}},\ \bibinfo
  {pages} {075001} (\bibinfo {year} {2016})},\ \Eprint
  {http://arxiv.org/abs/1603.04439} {arXiv:1603.04439 [hep-ph]} \BibitemShut
  {NoStop}%
\bibitem [{\citenamefont {Lee}\ and\ \citenamefont
  {Mohapatra}(2017)}]{Lee:2016wiy}%
  \BibitemOpen
  \bibfield  {author} {\bibinfo {author} {\bibfnamefont {C.-H.}\ \bibnamefont
  {Lee}}\ and\ \bibinfo {author} {\bibfnamefont {R.~N.}\ \bibnamefont
  {Mohapatra}},\ }\href {\doibase 10.1007/JHEP02(2017)080} {\bibfield
  {journal} {\bibinfo  {journal} {JHEP}\ }\textbf {\bibinfo {volume} {02}},\
  \bibinfo {pages} {080} (\bibinfo {year} {2017})},\ \Eprint
  {http://arxiv.org/abs/1611.05478} {arXiv:1611.05478 [hep-ph]} \BibitemShut
  {NoStop}%
\bibitem [{\citenamefont {Boucenna}\ and\ \citenamefont
  {Shafi}(2018)}]{Boucenna:2017fna}%
  \BibitemOpen
  \bibfield  {author} {\bibinfo {author} {\bibfnamefont {S.~M.}\ \bibnamefont
  {Boucenna}}\ and\ \bibinfo {author} {\bibfnamefont {Q.}~\bibnamefont
  {Shafi}},\ }\href {\doibase 10.1103/PhysRevD.97.075012} {\bibfield  {journal}
  {\bibinfo  {journal} {Phys. Rev. D}\ }\textbf {\bibinfo {volume} {97}},\
  \bibinfo {pages} {075012} (\bibinfo {year} {2018})},\ \Eprint
  {http://arxiv.org/abs/1712.06526} {arXiv:1712.06526 [hep-ph]} \BibitemShut
  {NoStop}%
\bibitem [{\citenamefont {Daido}\ \emph {et~al.}(2018)\citenamefont {Daido},
  \citenamefont {Takahashi},\ and\ \citenamefont {Yokozaki}}]{Daido:2018dmu}%
  \BibitemOpen
  \bibfield  {author} {\bibinfo {author} {\bibfnamefont {R.}~\bibnamefont
  {Daido}}, \bibinfo {author} {\bibfnamefont {F.}~\bibnamefont {Takahashi}}, \
  and\ \bibinfo {author} {\bibfnamefont {N.}~\bibnamefont {Yokozaki}},\ }\href
  {\doibase 10.1016/j.physletb.2018.03.039} {\bibfield  {journal} {\bibinfo
  {journal} {Phys. Lett. B}\ }\textbf {\bibinfo {volume} {780}},\ \bibinfo
  {pages} {538} (\bibinfo {year} {2018})},\ \Eprint
  {http://arxiv.org/abs/1801.10344} {arXiv:1801.10344 [hep-ph]} \BibitemShut
  {NoStop}%
\bibitem [{\citenamefont {Di~Luzio}\ \emph {et~al.}(2018)\citenamefont
  {Di~Luzio}, \citenamefont {Ringwald},\ and\ \citenamefont
  {Tamarit}}]{DiLuzio:2018gqe}%
  \BibitemOpen
  \bibfield  {author} {\bibinfo {author} {\bibfnamefont {L.}~\bibnamefont
  {Di~Luzio}}, \bibinfo {author} {\bibfnamefont {A.}~\bibnamefont {Ringwald}},
  \ and\ \bibinfo {author} {\bibfnamefont {C.}~\bibnamefont {Tamarit}},\ }\href
  {\doibase 10.1103/PhysRevD.98.095011} {\bibfield  {journal} {\bibinfo
  {journal} {Phys. Rev. D}\ }\textbf {\bibinfo {volume} {98}},\ \bibinfo
  {pages} {095011} (\bibinfo {year} {2018})},\ \Eprint
  {http://arxiv.org/abs/1807.09769} {arXiv:1807.09769 [hep-ph]} \BibitemShut
  {NoStop}%
\bibitem [{\citenamefont {Ernst}\ \emph {et~al.}(2019)\citenamefont {Ernst},
  \citenamefont {Di~Luzio}, \citenamefont {Ringwald},\ and\ \citenamefont
  {Tamarit}}]{Ernst:2018rod}%
  \BibitemOpen
  \bibfield  {author} {\bibinfo {author} {\bibfnamefont {A.}~\bibnamefont
  {Ernst}}, \bibinfo {author} {\bibfnamefont {L.}~\bibnamefont {Di~Luzio}},
  \bibinfo {author} {\bibfnamefont {A.}~\bibnamefont {Ringwald}}, \ and\
  \bibinfo {author} {\bibfnamefont {C.}~\bibnamefont {Tamarit}},\ }\href
  {\doibase 10.22323/1.347.0054} {\bibfield  {journal} {\bibinfo  {journal}
  {PoS}\ }\textbf {\bibinfo {volume} {CORFU2018}},\ \bibinfo {pages} {054}
  (\bibinfo {year} {2019})},\ \Eprint {http://arxiv.org/abs/1811.11860}
  {arXiv:1811.11860 [hep-ph]} \BibitemShut {NoStop}%
\bibitem [{\citenamefont {Fileviez~P\'erez}\ \emph {et~al.}(2019)\citenamefont
  {Fileviez~P\'erez}, \citenamefont {Murgui},\ and\ \citenamefont
  {Plascencia}}]{FileviezPerez:2019fku}%
  \BibitemOpen
  \bibfield  {author} {\bibinfo {author} {\bibfnamefont {P.}~\bibnamefont
  {Fileviez~P\'erez}}, \bibinfo {author} {\bibfnamefont {C.}~\bibnamefont
  {Murgui}}, \ and\ \bibinfo {author} {\bibfnamefont {A.~D.}\ \bibnamefont
  {Plascencia}},\ }\href {\doibase 10.1007/JHEP11(2019)093} {\bibfield
  {journal} {\bibinfo  {journal} {JHEP}\ }\textbf {\bibinfo {volume} {11}},\
  \bibinfo {pages} {093} (\bibinfo {year} {2019})},\ \Eprint
  {http://arxiv.org/abs/1908.01772} {arXiv:1908.01772 [hep-ph]} \BibitemShut
  {NoStop}%
\bibitem [{\citenamefont {Fileviez~P\'erez}\ \emph {et~al.}(2020)\citenamefont
  {Fileviez~P\'erez}, \citenamefont {Murgui},\ and\ \citenamefont
  {Plascencia}}]{FileviezPerez:2019ssf}%
  \BibitemOpen
  \bibfield  {author} {\bibinfo {author} {\bibfnamefont {P.}~\bibnamefont
  {Fileviez~P\'erez}}, \bibinfo {author} {\bibfnamefont {C.}~\bibnamefont
  {Murgui}}, \ and\ \bibinfo {author} {\bibfnamefont {A.~D.}\ \bibnamefont
  {Plascencia}},\ }\href {\doibase 10.1007/JHEP01(2020)091} {\bibfield
  {journal} {\bibinfo  {journal} {JHEP}\ }\textbf {\bibinfo {volume} {01}},\
  \bibinfo {pages} {091} (\bibinfo {year} {2020})},\ \Eprint
  {http://arxiv.org/abs/1911.05738} {arXiv:1911.05738 [hep-ph]} \BibitemShut
  {NoStop}%
\bibitem [{\citenamefont {Bajc}\ \emph {et~al.}(2006)\citenamefont {Bajc},
  \citenamefont {Melfo}, \citenamefont {Senjanovic},\ and\ \citenamefont
  {Vissani}}]{Bajc:2005zf}%
  \BibitemOpen
  \bibfield  {author} {\bibinfo {author} {\bibfnamefont {B.}~\bibnamefont
  {Bajc}}, \bibinfo {author} {\bibfnamefont {A.}~\bibnamefont {Melfo}},
  \bibinfo {author} {\bibfnamefont {G.}~\bibnamefont {Senjanovic}}, \ and\
  \bibinfo {author} {\bibfnamefont {F.}~\bibnamefont {Vissani}},\ }\href
  {\doibase 10.1103/PhysRevD.73.055001} {\bibfield  {journal} {\bibinfo
  {journal} {Phys. Rev. D}\ }\textbf {\bibinfo {volume} {73}},\ \bibinfo
  {pages} {055001} (\bibinfo {year} {2006})},\ \Eprint
  {http://arxiv.org/abs/hep-ph/0510139} {arXiv:hep-ph/0510139} \BibitemShut
  {NoStop}%
\bibitem [{\citenamefont {Altarelli}\ and\ \citenamefont
  {Meloni}(2013)}]{Altarelli:2013aqa}%
  \BibitemOpen
  \bibfield  {author} {\bibinfo {author} {\bibfnamefont {G.}~\bibnamefont
  {Altarelli}}\ and\ \bibinfo {author} {\bibfnamefont {D.}~\bibnamefont
  {Meloni}},\ }\href {\doibase 10.1007/JHEP08(2013)021} {\bibfield  {journal}
  {\bibinfo  {journal} {JHEP}\ }\textbf {\bibinfo {volume} {08}},\ \bibinfo
  {pages} {021} (\bibinfo {year} {2013})},\ \Eprint
  {http://arxiv.org/abs/1305.1001} {arXiv:1305.1001 [hep-ph]} \BibitemShut
  {NoStop}%
\bibitem [{\citenamefont {Babu}\ and\ \citenamefont
  {Khan}(2015)}]{Babu:2015bna}%
  \BibitemOpen
  \bibfield  {author} {\bibinfo {author} {\bibfnamefont {K.~S.}\ \bibnamefont
  {Babu}}\ and\ \bibinfo {author} {\bibfnamefont {S.}~\bibnamefont {Khan}},\
  }\href {\doibase 10.1103/PhysRevD.92.075018} {\bibfield  {journal} {\bibinfo
  {journal} {Phys. Rev. D}\ }\textbf {\bibinfo {volume} {92}},\ \bibinfo
  {pages} {075018} (\bibinfo {year} {2015})},\ \Eprint
  {http://arxiv.org/abs/1507.06712} {arXiv:1507.06712 [hep-ph]} \BibitemShut
  {NoStop}%
\bibitem [{\citenamefont {Ernst}\ \emph {et~al.}(2018)\citenamefont {Ernst},
  \citenamefont {Ringwald},\ and\ \citenamefont {Tamarit}}]{Ernst:2018bib}%
  \BibitemOpen
  \bibfield  {author} {\bibinfo {author} {\bibfnamefont {A.}~\bibnamefont
  {Ernst}}, \bibinfo {author} {\bibfnamefont {A.}~\bibnamefont {Ringwald}}, \
  and\ \bibinfo {author} {\bibfnamefont {C.}~\bibnamefont {Tamarit}},\ }\href
  {\doibase 10.1007/JHEP02(2018)103} {\bibfield  {journal} {\bibinfo  {journal}
  {JHEP}\ }\textbf {\bibinfo {volume} {02}},\ \bibinfo {pages} {103} (\bibinfo
  {year} {2018})},\ \Eprint {http://arxiv.org/abs/1801.04906} {arXiv:1801.04906
  [hep-ph]} \BibitemShut {NoStop}%
\bibitem [{\citenamefont {Corian\`o}\ \emph {et~al.}(2020)\citenamefont
  {Corian\`o}, \citenamefont {Frampton}, \citenamefont {Tatullo},\ and\
  \citenamefont {Theofilopoulos}}]{Coriano:2019vjl}%
  \BibitemOpen
  \bibfield  {author} {\bibinfo {author} {\bibfnamefont {C.}~\bibnamefont
  {Corian\`o}}, \bibinfo {author} {\bibfnamefont {P.~H.}\ \bibnamefont
  {Frampton}}, \bibinfo {author} {\bibfnamefont {A.}~\bibnamefont {Tatullo}}, \
  and\ \bibinfo {author} {\bibfnamefont {D.}~\bibnamefont {Theofilopoulos}},\
  }\href {\doibase 10.1016/j.physletb.2020.135273} {\bibfield  {journal}
  {\bibinfo  {journal} {Phys. Lett. B}\ }\textbf {\bibinfo {volume} {802}},\
  \bibinfo {pages} {135273} (\bibinfo {year} {2020})},\ \Eprint
  {http://arxiv.org/abs/1906.05810} {arXiv:1906.05810 [hep-ph]} \BibitemShut
  {NoStop}%
\bibitem [{\citenamefont {Di~Luzio}(2020)}]{DiLuzio:2020qio}%
  \BibitemOpen
  \bibfield  {author} {\bibinfo {author} {\bibfnamefont {L.}~\bibnamefont
  {Di~Luzio}},\ }\href {\doibase 10.1007/JHEP11(2020)074} {\bibfield  {journal}
  {\bibinfo  {journal} {JHEP}\ }\textbf {\bibinfo {volume} {11}},\ \bibinfo
  {pages} {074} (\bibinfo {year} {2020})},\ \Eprint
  {http://arxiv.org/abs/2008.09119} {arXiv:2008.09119 [hep-ph]} \BibitemShut
  {NoStop}%
\bibitem [{\citenamefont {Coriano}\ and\ \citenamefont
  {Frampton}(2018)}]{Coriano:2017ghp}%
  \BibitemOpen
  \bibfield  {author} {\bibinfo {author} {\bibfnamefont {C.}~\bibnamefont
  {Coriano}}\ and\ \bibinfo {author} {\bibfnamefont {P.~H.}\ \bibnamefont
  {Frampton}},\ }\href {\doibase 10.1016/j.physletb.2018.05.067} {\bibfield
  {journal} {\bibinfo  {journal} {Phys. Lett. B}\ }\textbf {\bibinfo {volume}
  {782}},\ \bibinfo {pages} {380} (\bibinfo {year} {2018})},\ \Eprint
  {http://arxiv.org/abs/1712.03865} {arXiv:1712.03865 [hep-ph]} \BibitemShut
  {NoStop}%
\bibitem [{\citenamefont {Dine}\ and\ \citenamefont
  {Seiberg}(1986)}]{Dine:1986bg}%
  \BibitemOpen
  \bibfield  {author} {\bibinfo {author} {\bibfnamefont {M.}~\bibnamefont
  {Dine}}\ and\ \bibinfo {author} {\bibfnamefont {N.}~\bibnamefont {Seiberg}},\
  }\href {\doibase 10.1016/0550-3213(86)90043-X} {\bibfield  {journal}
  {\bibinfo  {journal} {Nucl. Phys. B}\ }\textbf {\bibinfo {volume} {273}},\
  \bibinfo {pages} {109} (\bibinfo {year} {1986})}\BibitemShut {NoStop}%
\bibitem [{\citenamefont {Barr}\ and\ \citenamefont
  {Seckel}(1992)}]{Barr:1992qq}%
  \BibitemOpen
  \bibfield  {author} {\bibinfo {author} {\bibfnamefont {S.~M.}\ \bibnamefont
  {Barr}}\ and\ \bibinfo {author} {\bibfnamefont {D.}~\bibnamefont {Seckel}},\
  }\href {\doibase 10.1103/PhysRevD.46.539} {\bibfield  {journal} {\bibinfo
  {journal} {Phys. Rev. D}\ }\textbf {\bibinfo {volume} {46}},\ \bibinfo
  {pages} {539} (\bibinfo {year} {1992})}\BibitemShut {NoStop}%
\bibitem [{\citenamefont {Kamionkowski}\ and\ \citenamefont
  {March-Russell}(1992)}]{Kamionkowski:1992mf}%
  \BibitemOpen
  \bibfield  {author} {\bibinfo {author} {\bibfnamefont {M.}~\bibnamefont
  {Kamionkowski}}\ and\ \bibinfo {author} {\bibfnamefont {J.}~\bibnamefont
  {March-Russell}},\ }\href {\doibase 10.1016/0370-2693(92)90492-M} {\bibfield
  {journal} {\bibinfo  {journal} {Phys. Lett. B}\ }\textbf {\bibinfo {volume}
  {282}},\ \bibinfo {pages} {137} (\bibinfo {year} {1992})},\ \Eprint
  {http://arxiv.org/abs/hep-th/9202003} {arXiv:hep-th/9202003} \BibitemShut
  {NoStop}%
\bibitem [{\citenamefont {Holman}\ \emph {et~al.}(1992)\citenamefont {Holman},
  \citenamefont {Hsu}, \citenamefont {Kephart}, \citenamefont {Kolb},
  \citenamefont {Watkins},\ and\ \citenamefont {Widrow}}]{Holman:1992us}%
  \BibitemOpen
  \bibfield  {author} {\bibinfo {author} {\bibfnamefont {R.}~\bibnamefont
  {Holman}}, \bibinfo {author} {\bibfnamefont {S.~D.~H.}\ \bibnamefont {Hsu}},
  \bibinfo {author} {\bibfnamefont {T.~W.}\ \bibnamefont {Kephart}}, \bibinfo
  {author} {\bibfnamefont {E.~W.}\ \bibnamefont {Kolb}}, \bibinfo {author}
  {\bibfnamefont {R.}~\bibnamefont {Watkins}}, \ and\ \bibinfo {author}
  {\bibfnamefont {L.~M.}\ \bibnamefont {Widrow}},\ }\href {\doibase
  10.1016/0370-2693(92)90491-L} {\bibfield  {journal} {\bibinfo  {journal}
  {Phys. Lett. B}\ }\textbf {\bibinfo {volume} {282}},\ \bibinfo {pages} {132}
  (\bibinfo {year} {1992})},\ \Eprint {http://arxiv.org/abs/hep-ph/9203206}
  {arXiv:hep-ph/9203206} \BibitemShut {NoStop}%
\bibitem [{\citenamefont {Ghigna}\ \emph {et~al.}(1992)\citenamefont {Ghigna},
  \citenamefont {Lusignoli},\ and\ \citenamefont {Roncadelli}}]{Ghigna:1992iv}%
  \BibitemOpen
  \bibfield  {author} {\bibinfo {author} {\bibfnamefont {S.}~\bibnamefont
  {Ghigna}}, \bibinfo {author} {\bibfnamefont {M.}~\bibnamefont {Lusignoli}}, \
  and\ \bibinfo {author} {\bibfnamefont {M.}~\bibnamefont {Roncadelli}},\
  }\href {\doibase 10.1016/0370-2693(92)90019-Z} {\bibfield  {journal}
  {\bibinfo  {journal} {Phys. Lett. B}\ }\textbf {\bibinfo {volume} {283}},\
  \bibinfo {pages} {278} (\bibinfo {year} {1992})}\BibitemShut {NoStop}%
\bibitem [{\citenamefont {Carpenter}\ \emph {et~al.}(2009)\citenamefont
  {Carpenter}, \citenamefont {Dine},\ and\ \citenamefont
  {Festuccia}}]{Carpenter:2009zs}%
  \BibitemOpen
  \bibfield  {author} {\bibinfo {author} {\bibfnamefont {L.~M.}\ \bibnamefont
  {Carpenter}}, \bibinfo {author} {\bibfnamefont {M.}~\bibnamefont {Dine}}, \
  and\ \bibinfo {author} {\bibfnamefont {G.}~\bibnamefont {Festuccia}},\ }\href
  {\doibase 10.1103/PhysRevD.80.125017} {\bibfield  {journal} {\bibinfo
  {journal} {Phys. Rev. D}\ }\textbf {\bibinfo {volume} {80}},\ \bibinfo
  {pages} {125017} (\bibinfo {year} {2009})},\ \Eprint
  {http://arxiv.org/abs/0906.1273} {arXiv:0906.1273 [hep-th]} \BibitemShut
  {NoStop}%
\bibitem [{\citenamefont {Harigaya}\ \emph {et~al.}(2013)\citenamefont
  {Harigaya}, \citenamefont {Ibe}, \citenamefont {Schmitz},\ and\ \citenamefont
  {Yanagida}}]{Harigaya:2013vja}%
  \BibitemOpen
  \bibfield  {author} {\bibinfo {author} {\bibfnamefont {K.}~\bibnamefont
  {Harigaya}}, \bibinfo {author} {\bibfnamefont {M.}~\bibnamefont {Ibe}},
  \bibinfo {author} {\bibfnamefont {K.}~\bibnamefont {Schmitz}}, \ and\
  \bibinfo {author} {\bibfnamefont {T.~T.}\ \bibnamefont {Yanagida}},\ }\href
  {\doibase 10.1103/PhysRevD.88.075022} {\bibfield  {journal} {\bibinfo
  {journal} {Phys. Rev. D}\ }\textbf {\bibinfo {volume} {88}},\ \bibinfo
  {pages} {075022} (\bibinfo {year} {2013})},\ \Eprint
  {http://arxiv.org/abs/1308.1227} {arXiv:1308.1227 [hep-ph]} \BibitemShut
  {NoStop}%
\bibitem [{\citenamefont {Dias}\ \emph {et~al.}(2014)\citenamefont {Dias},
  \citenamefont {Machado}, \citenamefont {Nishi}, \citenamefont {Ringwald},\
  and\ \citenamefont {Vaudrevange}}]{Dias:2014osa}%
  \BibitemOpen
  \bibfield  {author} {\bibinfo {author} {\bibfnamefont {A.~G.}\ \bibnamefont
  {Dias}}, \bibinfo {author} {\bibfnamefont {A.~C.~B.}\ \bibnamefont
  {Machado}}, \bibinfo {author} {\bibfnamefont {C.~C.}\ \bibnamefont {Nishi}},
  \bibinfo {author} {\bibfnamefont {A.}~\bibnamefont {Ringwald}}, \ and\
  \bibinfo {author} {\bibfnamefont {P.}~\bibnamefont {Vaudrevange}},\ }\href
  {\doibase 10.1007/JHEP06(2014)037} {\bibfield  {journal} {\bibinfo  {journal}
  {JHEP}\ }\textbf {\bibinfo {volume} {06}},\ \bibinfo {pages} {037} (\bibinfo
  {year} {2014})},\ \Eprint {http://arxiv.org/abs/1403.5760} {arXiv:1403.5760
  [hep-ph]} \BibitemShut {NoStop}%
\bibitem [{\citenamefont {Harigaya}\ \emph {et~al.}(2015)\citenamefont
  {Harigaya}, \citenamefont {Ibe}, \citenamefont {Schmitz},\ and\ \citenamefont
  {Yanagida}}]{Harigaya:2015soa}%
  \BibitemOpen
  \bibfield  {author} {\bibinfo {author} {\bibfnamefont {K.}~\bibnamefont
  {Harigaya}}, \bibinfo {author} {\bibfnamefont {M.}~\bibnamefont {Ibe}},
  \bibinfo {author} {\bibfnamefont {K.}~\bibnamefont {Schmitz}}, \ and\
  \bibinfo {author} {\bibfnamefont {T.~T.}\ \bibnamefont {Yanagida}},\ }\href
  {\doibase 10.1103/PhysRevD.92.075003} {\bibfield  {journal} {\bibinfo
  {journal} {Phys. Rev. D}\ }\textbf {\bibinfo {volume} {92}},\ \bibinfo
  {pages} {075003} (\bibinfo {year} {2015})},\ \Eprint
  {http://arxiv.org/abs/1505.07388} {arXiv:1505.07388 [hep-ph]} \BibitemShut
  {NoStop}%
\bibitem [{\citenamefont {Redi}\ and\ \citenamefont
  {Sato}(2016)}]{Redi:2016esr}%
  \BibitemOpen
  \bibfield  {author} {\bibinfo {author} {\bibfnamefont {M.}~\bibnamefont
  {Redi}}\ and\ \bibinfo {author} {\bibfnamefont {R.}~\bibnamefont {Sato}},\
  }\href {\doibase 10.1007/JHEP05(2016)104} {\bibfield  {journal} {\bibinfo
  {journal} {JHEP}\ }\textbf {\bibinfo {volume} {05}},\ \bibinfo {pages} {104}
  (\bibinfo {year} {2016})},\ \Eprint {http://arxiv.org/abs/1602.05427}
  {arXiv:1602.05427 [hep-ph]} \BibitemShut {NoStop}%
\bibitem [{\citenamefont {Higaki}\ \emph {et~al.}(2016)\citenamefont {Higaki},
  \citenamefont {Jeong}, \citenamefont {Kitajima},\ and\ \citenamefont
  {Takahashi}}]{Higaki:2016yqk}%
  \BibitemOpen
  \bibfield  {author} {\bibinfo {author} {\bibfnamefont {T.}~\bibnamefont
  {Higaki}}, \bibinfo {author} {\bibfnamefont {K.~S.}\ \bibnamefont {Jeong}},
  \bibinfo {author} {\bibfnamefont {N.}~\bibnamefont {Kitajima}}, \ and\
  \bibinfo {author} {\bibfnamefont {F.}~\bibnamefont {Takahashi}},\ }\href
  {\doibase 10.1007/JHEP06(2016)150} {\bibfield  {journal} {\bibinfo  {journal}
  {JHEP}\ }\textbf {\bibinfo {volume} {06}},\ \bibinfo {pages} {150} (\bibinfo
  {year} {2016})},\ \Eprint {http://arxiv.org/abs/1603.02090} {arXiv:1603.02090
  [hep-ph]} \BibitemShut {NoStop}%
\bibitem [{\citenamefont {Fukuda}\ \emph {et~al.}(2017)\citenamefont {Fukuda},
  \citenamefont {Ibe}, \citenamefont {Suzuki},\ and\ \citenamefont
  {Yanagida}}]{Fukuda:2017ylt}%
  \BibitemOpen
  \bibfield  {author} {\bibinfo {author} {\bibfnamefont {H.}~\bibnamefont
  {Fukuda}}, \bibinfo {author} {\bibfnamefont {M.}~\bibnamefont {Ibe}},
  \bibinfo {author} {\bibfnamefont {M.}~\bibnamefont {Suzuki}}, \ and\ \bibinfo
  {author} {\bibfnamefont {T.~T.}\ \bibnamefont {Yanagida}},\ }\href {\doibase
  10.1016/j.physletb.2017.05.071} {\bibfield  {journal} {\bibinfo  {journal}
  {Phys. Lett. B}\ }\textbf {\bibinfo {volume} {771}},\ \bibinfo {pages} {327}
  (\bibinfo {year} {2017})},\ \Eprint {http://arxiv.org/abs/1703.01112}
  {arXiv:1703.01112 [hep-ph]} \BibitemShut {NoStop}%
\bibitem [{\citenamefont {Di~Luzio}\ \emph
  {et~al.}(2017{\natexlab{a}})\citenamefont {Di~Luzio}, \citenamefont {Nardi},\
  and\ \citenamefont {Ubaldi}}]{DiLuzio:2017tjx}%
  \BibitemOpen
  \bibfield  {author} {\bibinfo {author} {\bibfnamefont {L.}~\bibnamefont
  {Di~Luzio}}, \bibinfo {author} {\bibfnamefont {E.}~\bibnamefont {Nardi}}, \
  and\ \bibinfo {author} {\bibfnamefont {L.}~\bibnamefont {Ubaldi}},\ }\href
  {\doibase 10.1103/PhysRevLett.119.011801} {\bibfield  {journal} {\bibinfo
  {journal} {Phys. Rev. Lett.}\ }\textbf {\bibinfo {volume} {119}},\ \bibinfo
  {pages} {011801} (\bibinfo {year} {2017}{\natexlab{a}})},\ \Eprint
  {http://arxiv.org/abs/1704.01122} {arXiv:1704.01122 [hep-ph]} \BibitemShut
  {NoStop}%
\bibitem [{\citenamefont {Duerr}\ \emph {et~al.}(2018)\citenamefont {Duerr},
  \citenamefont {Schmidt-Hoberg},\ and\ \citenamefont {Unwin}}]{Duerr:2017amf}%
  \BibitemOpen
  \bibfield  {author} {\bibinfo {author} {\bibfnamefont {M.}~\bibnamefont
  {Duerr}}, \bibinfo {author} {\bibfnamefont {K.}~\bibnamefont
  {Schmidt-Hoberg}}, \ and\ \bibinfo {author} {\bibfnamefont {J.}~\bibnamefont
  {Unwin}},\ }\href {\doibase 10.1016/j.physletb.2018.03.054} {\bibfield
  {journal} {\bibinfo  {journal} {Phys. Lett. B}\ }\textbf {\bibinfo {volume}
  {780}},\ \bibinfo {pages} {553} (\bibinfo {year} {2018})},\ \Eprint
  {http://arxiv.org/abs/1712.01841} {arXiv:1712.01841 [hep-ph]} \BibitemShut
  {NoStop}%
\bibitem [{\citenamefont {Lillard}\ and\ \citenamefont
  {Tait}(2018)}]{Lillard:2018fdt}%
  \BibitemOpen
  \bibfield  {author} {\bibinfo {author} {\bibfnamefont {B.}~\bibnamefont
  {Lillard}}\ and\ \bibinfo {author} {\bibfnamefont {T.~M.~P.}\ \bibnamefont
  {Tait}},\ }\href {\doibase 10.1007/JHEP11(2018)199} {\bibfield  {journal}
  {\bibinfo  {journal} {JHEP}\ }\textbf {\bibinfo {volume} {11}},\ \bibinfo
  {pages} {199} (\bibinfo {year} {2018})},\ \Eprint
  {http://arxiv.org/abs/1811.03089} {arXiv:1811.03089 [hep-ph]} \BibitemShut
  {NoStop}%
\bibitem [{\citenamefont {Lee}\ and\ \citenamefont {Yin}(2019)}]{Lee:2018yak}%
  \BibitemOpen
  \bibfield  {author} {\bibinfo {author} {\bibfnamefont {H.-S.}\ \bibnamefont
  {Lee}}\ and\ \bibinfo {author} {\bibfnamefont {W.}~\bibnamefont {Yin}},\
  }\href {\doibase 10.1103/PhysRevD.99.015041} {\bibfield  {journal} {\bibinfo
  {journal} {Phys. Rev. D}\ }\textbf {\bibinfo {volume} {99}},\ \bibinfo
  {pages} {015041} (\bibinfo {year} {2019})},\ \Eprint
  {http://arxiv.org/abs/1811.04039} {arXiv:1811.04039 [hep-ph]} \BibitemShut
  {NoStop}%
\bibitem [{\citenamefont {Gavela}\ \emph {et~al.}(2019)\citenamefont {Gavela},
  \citenamefont {Ibe}, \citenamefont {Quilez},\ and\ \citenamefont
  {Yanagida}}]{Gavela:2018paw}%
  \BibitemOpen
  \bibfield  {author} {\bibinfo {author} {\bibfnamefont {M.~B.}\ \bibnamefont
  {Gavela}}, \bibinfo {author} {\bibfnamefont {M.}~\bibnamefont {Ibe}},
  \bibinfo {author} {\bibfnamefont {P.}~\bibnamefont {Quilez}}, \ and\ \bibinfo
  {author} {\bibfnamefont {T.~T.}\ \bibnamefont {Yanagida}},\ }\href {\doibase
  10.1140/epjc/s10052-019-7046-3} {\bibfield  {journal} {\bibinfo  {journal}
  {Eur. Phys. J. C}\ }\textbf {\bibinfo {volume} {79}},\ \bibinfo {pages} {542}
  (\bibinfo {year} {2019})},\ \Eprint {http://arxiv.org/abs/1812.08174}
  {arXiv:1812.08174 [hep-ph]} \BibitemShut {NoStop}%
\bibitem [{\citenamefont {Ardu}\ \emph {et~al.}(2020)\citenamefont {Ardu},
  \citenamefont {Di~Luzio}, \citenamefont {Landini}, \citenamefont {Strumia},
  \citenamefont {Teresi},\ and\ \citenamefont {Wang}}]{Ardu:2020qmo}%
  \BibitemOpen
  \bibfield  {author} {\bibinfo {author} {\bibfnamefont {M.}~\bibnamefont
  {Ardu}}, \bibinfo {author} {\bibfnamefont {L.}~\bibnamefont {Di~Luzio}},
  \bibinfo {author} {\bibfnamefont {G.}~\bibnamefont {Landini}}, \bibinfo
  {author} {\bibfnamefont {A.}~\bibnamefont {Strumia}}, \bibinfo {author}
  {\bibfnamefont {D.}~\bibnamefont {Teresi}}, \ and\ \bibinfo {author}
  {\bibfnamefont {J.-W.}\ \bibnamefont {Wang}},\ }\href {\doibase
  10.1007/JHEP11(2020)090} {\bibfield  {journal} {\bibinfo  {journal} {JHEP}\
  }\textbf {\bibinfo {volume} {11}},\ \bibinfo {pages} {090} (\bibinfo {year}
  {2020})},\ \Eprint {http://arxiv.org/abs/2007.12663} {arXiv:2007.12663
  [hep-ph]} \BibitemShut {NoStop}%
\bibitem [{\citenamefont {Yin}(2020)}]{Yin:2020dfn}%
  \BibitemOpen
  \bibfield  {author} {\bibinfo {author} {\bibfnamefont {W.}~\bibnamefont
  {Yin}},\ }\href {\doibase 10.1007/JHEP10(2020)032} {\bibfield  {journal}
  {\bibinfo  {journal} {JHEP}\ }\textbf {\bibinfo {volume} {10}},\ \bibinfo
  {pages} {032} (\bibinfo {year} {2020})},\ \Eprint
  {http://arxiv.org/abs/2007.13320} {arXiv:2007.13320 [hep-ph]} \BibitemShut
  {NoStop}%
\bibitem [{\citenamefont {Nakai}\ and\ \citenamefont
  {Suzuki}(2021)}]{Nakai:2021nyf}%
  \BibitemOpen
  \bibfield  {author} {\bibinfo {author} {\bibfnamefont {Y.}~\bibnamefont
  {Nakai}}\ and\ \bibinfo {author} {\bibfnamefont {M.}~\bibnamefont {Suzuki}},\
  }\href {\doibase 10.1016/j.physletb.2021.136239} {\bibfield  {journal}
  {\bibinfo  {journal} {Phys. Lett. B}\ }\textbf {\bibinfo {volume} {816}},\
  \bibinfo {pages} {136239} (\bibinfo {year} {2021})},\ \Eprint
  {http://arxiv.org/abs/2102.01329} {arXiv:2102.01329 [hep-ph]} \BibitemShut
  {NoStop}%
\bibitem [{\citenamefont {Dimopoulos}\ \emph {et~al.}(1980)\citenamefont
  {Dimopoulos}, \citenamefont {Raby},\ and\ \citenamefont
  {Susskind}}]{Dimopoulos:1980hn}%
  \BibitemOpen
  \bibfield  {author} {\bibinfo {author} {\bibfnamefont {S.}~\bibnamefont
  {Dimopoulos}}, \bibinfo {author} {\bibfnamefont {S.}~\bibnamefont {Raby}}, \
  and\ \bibinfo {author} {\bibfnamefont {L.}~\bibnamefont {Susskind}},\ }\href
  {\doibase 10.1016/0550-3213(80)90215-1} {\bibfield  {journal} {\bibinfo
  {journal} {Nucl. Phys. B}\ }\textbf {\bibinfo {volume} {173}},\ \bibinfo
  {pages} {208} (\bibinfo {year} {1980})}\BibitemShut {NoStop}%
\bibitem [{\citenamefont {Abud}\ \emph {et~al.}(1977)\citenamefont {Abud},
  \citenamefont {Buccella}, \citenamefont {Ruegg},\ and\ \citenamefont
  {Savoy}}]{Abud:1977ej}%
  \BibitemOpen
  \bibfield  {author} {\bibinfo {author} {\bibfnamefont {M.}~\bibnamefont
  {Abud}}, \bibinfo {author} {\bibfnamefont {F.}~\bibnamefont {Buccella}},
  \bibinfo {author} {\bibfnamefont {H.}~\bibnamefont {Ruegg}}, \ and\ \bibinfo
  {author} {\bibfnamefont {C.~A.}\ \bibnamefont {Savoy}},\ }\href {\doibase
  10.1016/0370-2693(77)90380-X} {\bibfield  {journal} {\bibinfo  {journal}
  {Phys. Lett. B}\ }\textbf {\bibinfo {volume} {67}},\ \bibinfo {pages} {313}
  (\bibinfo {year} {1977})}\BibitemShut {NoStop}%
\bibitem [{\citenamefont {Langacker}\ \emph
  {et~al.}(1978{\natexlab{a}})\citenamefont {Langacker}, \citenamefont
  {Segre},\ and\ \citenamefont {Weldon}}]{Langacker:1977ai}%
  \BibitemOpen
  \bibfield  {author} {\bibinfo {author} {\bibfnamefont {P.}~\bibnamefont
  {Langacker}}, \bibinfo {author} {\bibfnamefont {G.}~\bibnamefont {Segre}}, \
  and\ \bibinfo {author} {\bibfnamefont {H.~A.}\ \bibnamefont {Weldon}},\
  }\href {\doibase 10.1016/0370-2693(78)90178-8} {\bibfield  {journal}
  {\bibinfo  {journal} {Phys. Lett. B}\ }\textbf {\bibinfo {volume} {73}},\
  \bibinfo {pages} {87} (\bibinfo {year} {1978}{\natexlab{a}})}\BibitemShut
  {NoStop}%
\bibitem [{\citenamefont {Inoue}\ \emph
  {et~al.}(1977{\natexlab{a}})\citenamefont {Inoue}, \citenamefont {Kakuto},\
  and\ \citenamefont {Nakano}}]{Inoue:1977qd}%
  \BibitemOpen
  \bibfield  {author} {\bibinfo {author} {\bibfnamefont {K.}~\bibnamefont
  {Inoue}}, \bibinfo {author} {\bibfnamefont {A.}~\bibnamefont {Kakuto}}, \
  and\ \bibinfo {author} {\bibfnamefont {Y.}~\bibnamefont {Nakano}},\ }\href
  {\doibase 10.1143/PTP.58.630} {\bibfield  {journal} {\bibinfo  {journal}
  {Prog. Theor. Phys.}\ }\textbf {\bibinfo {volume} {58}},\ \bibinfo {pages}
  {630} (\bibinfo {year} {1977}{\natexlab{a}})}\BibitemShut {NoStop}%
\bibitem [{\citenamefont {Inoue}\ \emph
  {et~al.}(1977{\natexlab{b}})\citenamefont {Inoue}, \citenamefont {Kakuto},
  \citenamefont {Komatsu},\ and\ \citenamefont {Nakano}}]{Inoue:1977an}%
  \BibitemOpen
  \bibfield  {author} {\bibinfo {author} {\bibfnamefont {K.}~\bibnamefont
  {Inoue}}, \bibinfo {author} {\bibfnamefont {A.}~\bibnamefont {Kakuto}},
  \bibinfo {author} {\bibfnamefont {H.}~\bibnamefont {Komatsu}}, \ and\
  \bibinfo {author} {\bibfnamefont {Y.}~\bibnamefont {Nakano}},\ }\href
  {\doibase 10.1143/PTP.58.1901} {\bibfield  {journal} {\bibinfo  {journal}
  {Prog. Theor. Phys.}\ }\textbf {\bibinfo {volume} {58}},\ \bibinfo {pages}
  {1901} (\bibinfo {year} {1977}{\natexlab{b}})}\BibitemShut {NoStop}%
\bibitem [{\citenamefont {Inoue}\ \emph
  {et~al.}(1977{\natexlab{c}})\citenamefont {Inoue}, \citenamefont {Kakuto},
  \citenamefont {Komatsu},\ and\ \citenamefont {Nakano}}]{Inoue:1977qw}%
  \BibitemOpen
  \bibfield  {author} {\bibinfo {author} {\bibfnamefont {K.}~\bibnamefont
  {Inoue}}, \bibinfo {author} {\bibfnamefont {A.}~\bibnamefont {Kakuto}},
  \bibinfo {author} {\bibfnamefont {H.}~\bibnamefont {Komatsu}}, \ and\
  \bibinfo {author} {\bibfnamefont {Y.}~\bibnamefont {Nakano}},\ }\href
  {\doibase 10.1143/PTP.58.1914} {\bibfield  {journal} {\bibinfo  {journal}
  {Prog. Theor. Phys.}\ }\textbf {\bibinfo {volume} {58}},\ \bibinfo {pages}
  {1914} (\bibinfo {year} {1977}{\natexlab{c}})}\BibitemShut {NoStop}%
\bibitem [{\citenamefont {Yun}(1978)}]{Yun:1978aa}%
  \BibitemOpen
  \bibfield  {author} {\bibinfo {author} {\bibfnamefont {S.~K.}\ \bibnamefont
  {Yun}},\ }\href {\doibase 10.1103/PhysRevD.18.3472} {\bibfield  {journal}
  {\bibinfo  {journal} {Phys. Rev. D}\ }\textbf {\bibinfo {volume} {18}},\
  \bibinfo {pages} {3472} (\bibinfo {year} {1978})}\BibitemShut {NoStop}%
\bibitem [{\citenamefont {Langacker}\ \emph
  {et~al.}(1978{\natexlab{b}})\citenamefont {Langacker}, \citenamefont
  {Segre},\ and\ \citenamefont {Weldon}}]{Langacker:1978fn}%
  \BibitemOpen
  \bibfield  {author} {\bibinfo {author} {\bibfnamefont {P.}~\bibnamefont
  {Langacker}}, \bibinfo {author} {\bibfnamefont {G.}~\bibnamefont {Segre}}, \
  and\ \bibinfo {author} {\bibfnamefont {H.~A.}\ \bibnamefont {Weldon}},\
  }\href {\doibase 10.1103/PhysRevD.18.552} {\bibfield  {journal} {\bibinfo
  {journal} {Phys. Rev. D}\ }\textbf {\bibinfo {volume} {18}},\ \bibinfo
  {pages} {552} (\bibinfo {year} {1978}{\natexlab{b}})}\BibitemShut {NoStop}%
\bibitem [{\citenamefont {Tomozawa}(1979)}]{Tomozawa:1978uz}%
  \BibitemOpen
  \bibfield  {author} {\bibinfo {author} {\bibfnamefont {Y.}~\bibnamefont
  {Tomozawa}},\ }\href {\doibase 10.1103/PhysRevD.19.1626} {\bibfield
  {journal} {\bibinfo  {journal} {Phys. Rev. D}\ }\textbf {\bibinfo {volume}
  {19}},\ \bibinfo {pages} {1626} (\bibinfo {year} {1979})},\ \bibinfo {note}
  {[Erratum: Phys.Rev.D 24, 1445 (1981)]}\BibitemShut {NoStop}%
\bibitem [{\citenamefont {Georgi}\ and\ \citenamefont
  {Pais}(1979)}]{Georgi:1978bv}%
  \BibitemOpen
  \bibfield  {author} {\bibinfo {author} {\bibfnamefont {H.}~\bibnamefont
  {Georgi}}\ and\ \bibinfo {author} {\bibfnamefont {A.}~\bibnamefont {Pais}},\
  }\href {\doibase 10.1103/PhysRevD.19.2746} {\bibfield  {journal} {\bibinfo
  {journal} {Phys. Rev. D}\ }\textbf {\bibinfo {volume} {19}},\ \bibinfo
  {pages} {2746} (\bibinfo {year} {1979})}\BibitemShut {NoStop}%
\bibitem [{\citenamefont {Kim}(1981)}]{Kim:1981jw}%
  \BibitemOpen
  \bibfield  {author} {\bibinfo {author} {\bibfnamefont {J.~E.}\ \bibnamefont
  {Kim}},\ }\href {\doibase 10.1016/0370-2693(81)91149-7} {\bibfield  {journal}
  {\bibinfo  {journal} {Phys. Lett. B}\ }\textbf {\bibinfo {volume} {107}},\
  \bibinfo {pages} {69} (\bibinfo {year} {1981})}\BibitemShut {NoStop}%
\bibitem [{\citenamefont {Fukugita}\ \emph {et~al.}(1982)\citenamefont
  {Fukugita}, \citenamefont {Yanagida},\ and\ \citenamefont
  {Yoshimura}}]{Fukugita:1981gn}%
  \BibitemOpen
  \bibfield  {author} {\bibinfo {author} {\bibfnamefont {M.}~\bibnamefont
  {Fukugita}}, \bibinfo {author} {\bibfnamefont {T.}~\bibnamefont {Yanagida}},
  \ and\ \bibinfo {author} {\bibfnamefont {M.}~\bibnamefont {Yoshimura}},\
  }\href {\doibase 10.1016/0370-2693(82)91092-9} {\bibfield  {journal}
  {\bibinfo  {journal} {Phys. Lett. B}\ }\textbf {\bibinfo {volume} {109}},\
  \bibinfo {pages} {369} (\bibinfo {year} {1982})}\BibitemShut {NoStop}%
\bibitem [{\citenamefont {Tabata}\ \emph {et~al.}(1984)\citenamefont {Tabata},
  \citenamefont {Umemura},\ and\ \citenamefont {Yamamoto}}]{Tabata:1983cr}%
  \BibitemOpen
  \bibfield  {author} {\bibinfo {author} {\bibfnamefont {K.}~\bibnamefont
  {Tabata}}, \bibinfo {author} {\bibfnamefont {I.}~\bibnamefont {Umemura}}, \
  and\ \bibinfo {author} {\bibfnamefont {K.}~\bibnamefont {Yamamoto}},\ }\href
  {\doibase 10.1143/PTP.71.615} {\bibfield  {journal} {\bibinfo  {journal}
  {Prog. Theor. Phys.}\ }\textbf {\bibinfo {volume} {71}},\ \bibinfo {pages}
  {615} (\bibinfo {year} {1984})}\BibitemShut {NoStop}%
\bibitem [{\citenamefont {Sen}(1985)}]{Sen:1983xj}%
  \BibitemOpen
  \bibfield  {author} {\bibinfo {author} {\bibfnamefont {A.}~\bibnamefont
  {Sen}},\ }\href {\doibase 10.1103/PhysRevD.31.900} {\bibfield  {journal}
  {\bibinfo  {journal} {Phys. Rev. D}\ }\textbf {\bibinfo {volume} {31}},\
  \bibinfo {pages} {900} (\bibinfo {year} {1985})}\BibitemShut {NoStop}%
\bibitem [{\citenamefont {Barr}(2012)}]{Barr:2011cz}%
  \BibitemOpen
  \bibfield  {author} {\bibinfo {author} {\bibfnamefont {S.~M.}\ \bibnamefont
  {Barr}},\ }\href {\doibase 10.1103/PhysRevD.85.013001} {\bibfield  {journal}
  {\bibinfo  {journal} {Phys. Rev. D}\ }\textbf {\bibinfo {volume} {85}},\
  \bibinfo {pages} {013001} (\bibinfo {year} {2012})},\ \Eprint
  {http://arxiv.org/abs/1109.2562} {arXiv:1109.2562 [hep-ph]} \BibitemShut
  {NoStop}%
\bibitem [{\citenamefont {Deppisch}\ \emph {et~al.}(2016)\citenamefont
  {Deppisch}, \citenamefont {Hati}, \citenamefont {Patra}, \citenamefont
  {Sarkar},\ and\ \citenamefont {Valle}}]{Deppisch:2016jzl}%
  \BibitemOpen
  \bibfield  {author} {\bibinfo {author} {\bibfnamefont {F.~F.}\ \bibnamefont
  {Deppisch}}, \bibinfo {author} {\bibfnamefont {C.}~\bibnamefont {Hati}},
  \bibinfo {author} {\bibfnamefont {S.}~\bibnamefont {Patra}}, \bibinfo
  {author} {\bibfnamefont {U.}~\bibnamefont {Sarkar}}, \ and\ \bibinfo {author}
  {\bibfnamefont {J.~W.~F.}\ \bibnamefont {Valle}},\ }\href {\doibase
  10.1016/j.physletb.2016.10.002} {\bibfield  {journal} {\bibinfo  {journal}
  {Phys. Lett. B}\ }\textbf {\bibinfo {volume} {762}},\ \bibinfo {pages} {432}
  (\bibinfo {year} {2016})},\ \Eprint {http://arxiv.org/abs/1608.05334}
  {arXiv:1608.05334 [hep-ph]} \BibitemShut {NoStop}%
\bibitem [{\citenamefont {Li}\ \emph {et~al.}(2020)\citenamefont {Li},
  \citenamefont {Pei}, \citenamefont {Xu},\ and\ \citenamefont
  {Zhang}}]{Li:2019qxy}%
  \BibitemOpen
  \bibfield  {author} {\bibinfo {author} {\bibfnamefont {T.}~\bibnamefont
  {Li}}, \bibinfo {author} {\bibfnamefont {J.}~\bibnamefont {Pei}}, \bibinfo
  {author} {\bibfnamefont {F.}~\bibnamefont {Xu}}, \ and\ \bibinfo {author}
  {\bibfnamefont {W.}~\bibnamefont {Zhang}},\ }\href {\doibase
  10.1103/PhysRevD.102.016004} {\bibfield  {journal} {\bibinfo  {journal}
  {Phys. Rev. D}\ }\textbf {\bibinfo {volume} {102}},\ \bibinfo {pages}
  {016004} (\bibinfo {year} {2020})},\ \Eprint
  {http://arxiv.org/abs/1911.09551} {arXiv:1911.09551 [hep-ph]} \BibitemShut
  {NoStop}%
\bibitem [{\citenamefont {Chacko}\ \emph {et~al.}(2020)\citenamefont {Chacko},
  \citenamefont {Dev}, \citenamefont {Mohapatra},\ and\ \citenamefont
  {Thapa}}]{Chacko:2020tbu}%
  \BibitemOpen
  \bibfield  {author} {\bibinfo {author} {\bibfnamefont {Z.}~\bibnamefont
  {Chacko}}, \bibinfo {author} {\bibfnamefont {P.~S.~B.}\ \bibnamefont {Dev}},
  \bibinfo {author} {\bibfnamefont {R.~N.}\ \bibnamefont {Mohapatra}}, \ and\
  \bibinfo {author} {\bibfnamefont {A.}~\bibnamefont {Thapa}},\ }\href
  {\doibase 10.1103/PhysRevD.102.035020} {\bibfield  {journal} {\bibinfo
  {journal} {Phys. Rev. D}\ }\textbf {\bibinfo {volume} {102}},\ \bibinfo
  {pages} {035020} (\bibinfo {year} {2020})},\ \Eprint
  {http://arxiv.org/abs/2005.05413} {arXiv:2005.05413 [hep-ph]} \BibitemShut
  {NoStop}%
\bibitem [{\citenamefont {Angelescu}\ \emph {et~al.}(2021)\citenamefont
  {Angelescu}, \citenamefont {Bally}, \citenamefont {Blasi},\ and\
  \citenamefont {Goertz}}]{Angelescu:2021nbp}%
  \BibitemOpen
  \bibfield  {author} {\bibinfo {author} {\bibfnamefont {A.}~\bibnamefont
  {Angelescu}}, \bibinfo {author} {\bibfnamefont {A.}~\bibnamefont {Bally}},
  \bibinfo {author} {\bibfnamefont {S.}~\bibnamefont {Blasi}}, \ and\ \bibinfo
  {author} {\bibfnamefont {F.}~\bibnamefont {Goertz}},\ }\href@noop {} {\
  (\bibinfo {year} {2021})},\ \Eprint {http://arxiv.org/abs/2104.07366}
  {arXiv:2104.07366 [hep-ph]} \BibitemShut {NoStop}%
\bibitem [{\citenamefont {Berezhiani}\ and\ \citenamefont
  {Dvali}(1989)}]{Berezhiani:1989bd}%
  \BibitemOpen
  \bibfield  {author} {\bibinfo {author} {\bibfnamefont {Z.~G.}\ \bibnamefont
  {Berezhiani}}\ and\ \bibinfo {author} {\bibfnamefont {G.~R.}\ \bibnamefont
  {Dvali}},\ }\href@noop {} {\bibfield  {journal} {\bibinfo  {journal} {Bull.
  Lebedev Phys. Inst.}\ }\textbf {\bibinfo {volume} {5}},\ \bibinfo {pages}
  {55} (\bibinfo {year} {1989})}\BibitemShut {NoStop}%
\bibitem [{\citenamefont {Dvali}(1994)}]{Dvali:1993yf}%
  \BibitemOpen
  \bibfield  {author} {\bibinfo {author} {\bibfnamefont {G.~R.}\ \bibnamefont
  {Dvali}},\ }\href {\doibase 10.1016/0370-2693(94)00075-1} {\bibfield
  {journal} {\bibinfo  {journal} {Phys. Lett. B}\ }\textbf {\bibinfo {volume}
  {324}},\ \bibinfo {pages} {59} (\bibinfo {year} {1994})}\BibitemShut
  {NoStop}%
\bibitem [{\citenamefont {Berezhiani}(1995)}]{Berezhiani:1995dt}%
  \BibitemOpen
  \bibfield  {author} {\bibinfo {author} {\bibfnamefont {Z.}~\bibnamefont
  {Berezhiani}},\ }\href {\doibase 10.1016/0370-2693(95)00705-P} {\bibfield
  {journal} {\bibinfo  {journal} {Phys. Lett. B}\ }\textbf {\bibinfo {volume}
  {355}},\ \bibinfo {pages} {481} (\bibinfo {year} {1995})},\ \Eprint
  {http://arxiv.org/abs/hep-ph/9503366} {arXiv:hep-ph/9503366} \BibitemShut
  {NoStop}%
\bibitem [{\citenamefont {Berezhiani}\ \emph {et~al.}(1995)\citenamefont
  {Berezhiani}, \citenamefont {Csaki},\ and\ \citenamefont
  {Randall}}]{Berezhiani:1995sb}%
  \BibitemOpen
  \bibfield  {author} {\bibinfo {author} {\bibfnamefont {Z.}~\bibnamefont
  {Berezhiani}}, \bibinfo {author} {\bibfnamefont {C.}~\bibnamefont {Csaki}}, \
  and\ \bibinfo {author} {\bibfnamefont {L.}~\bibnamefont {Randall}},\ }\href
  {\doibase 10.1016/0550-3213(95)00183-S} {\bibfield  {journal} {\bibinfo
  {journal} {Nucl. Phys. B}\ }\textbf {\bibinfo {volume} {444}},\ \bibinfo
  {pages} {61} (\bibinfo {year} {1995})},\ \Eprint
  {http://arxiv.org/abs/hep-ph/9501336} {arXiv:hep-ph/9501336} \BibitemShut
  {NoStop}%
\bibitem [{\citenamefont {Barr}(1998)}]{Barr:1997pt}%
  \BibitemOpen
  \bibfield  {author} {\bibinfo {author} {\bibfnamefont {S.~M.}\ \bibnamefont
  {Barr}},\ }\href {\doibase 10.1103/PhysRevD.57.190} {\bibfield  {journal}
  {\bibinfo  {journal} {Phys. Rev. D}\ }\textbf {\bibinfo {volume} {57}},\
  \bibinfo {pages} {190} (\bibinfo {year} {1998})},\ \Eprint
  {http://arxiv.org/abs/hep-ph/9705266} {arXiv:hep-ph/9705266} \BibitemShut
  {NoStop}%
\bibitem [{\citenamefont {Chacko}\ and\ \citenamefont
  {Mohapatra}(1998)}]{Chacko:1998zz}%
  \BibitemOpen
  \bibfield  {author} {\bibinfo {author} {\bibfnamefont {Z.}~\bibnamefont
  {Chacko}}\ and\ \bibinfo {author} {\bibfnamefont {R.~N.}\ \bibnamefont
  {Mohapatra}},\ }\href {\doibase 10.1016/S0370-2693(98)01263-5} {\bibfield
  {journal} {\bibinfo  {journal} {Phys. Lett. B}\ }\textbf {\bibinfo {volume}
  {442}},\ \bibinfo {pages} {199} (\bibinfo {year} {1998})},\ \Eprint
  {http://arxiv.org/abs/hep-ph/9809345} {arXiv:hep-ph/9809345} \BibitemShut
  {NoStop}%
\bibitem [{\citenamefont {Pisano}\ and\ \citenamefont
  {Pleitez}(1992)}]{Pisano:1991ee}%
  \BibitemOpen
  \bibfield  {author} {\bibinfo {author} {\bibfnamefont {F.}~\bibnamefont
  {Pisano}}\ and\ \bibinfo {author} {\bibfnamefont {V.}~\bibnamefont
  {Pleitez}},\ }\href {\doibase 10.1103/PhysRevD.46.410} {\bibfield  {journal}
  {\bibinfo  {journal} {Phys. Rev. D}\ }\textbf {\bibinfo {volume} {46}},\
  \bibinfo {pages} {410} (\bibinfo {year} {1992})},\ \Eprint
  {http://arxiv.org/abs/hep-ph/9206242} {arXiv:hep-ph/9206242} \BibitemShut
  {NoStop}%
\bibitem [{\citenamefont {Foot}\ \emph {et~al.}(1993)\citenamefont {Foot},
  \citenamefont {Hernandez}, \citenamefont {Pisano},\ and\ \citenamefont
  {Pleitez}}]{Foot:1992rh}%
  \BibitemOpen
  \bibfield  {author} {\bibinfo {author} {\bibfnamefont {R.}~\bibnamefont
  {Foot}}, \bibinfo {author} {\bibfnamefont {O.~F.}\ \bibnamefont {Hernandez}},
  \bibinfo {author} {\bibfnamefont {F.}~\bibnamefont {Pisano}}, \ and\ \bibinfo
  {author} {\bibfnamefont {V.}~\bibnamefont {Pleitez}},\ }\href {\doibase
  10.1103/PhysRevD.47.4158} {\bibfield  {journal} {\bibinfo  {journal} {Phys.
  Rev. D}\ }\textbf {\bibinfo {volume} {47}},\ \bibinfo {pages} {4158}
  (\bibinfo {year} {1993})},\ \Eprint {http://arxiv.org/abs/hep-ph/9207264}
  {arXiv:hep-ph/9207264} \BibitemShut {NoStop}%
\bibitem [{\citenamefont {Ng}(1994)}]{Ng:1992st}%
  \BibitemOpen
  \bibfield  {author} {\bibinfo {author} {\bibfnamefont {D.}~\bibnamefont
  {Ng}},\ }\href {\doibase 10.1103/PhysRevD.49.4805} {\bibfield  {journal}
  {\bibinfo  {journal} {Phys. Rev. D}\ }\textbf {\bibinfo {volume} {49}},\
  \bibinfo {pages} {4805} (\bibinfo {year} {1994})},\ \Eprint
  {http://arxiv.org/abs/hep-ph/9212284} {arXiv:hep-ph/9212284} \BibitemShut
  {NoStop}%
\bibitem [{\citenamefont {Tonasse}(1996)}]{Tonasse:1996cx}%
  \BibitemOpen
  \bibfield  {author} {\bibinfo {author} {\bibfnamefont {M.~D.}\ \bibnamefont
  {Tonasse}},\ }\href {\doibase 10.1016/0370-2693(96)00481-9} {\bibfield
  {journal} {\bibinfo  {journal} {Phys. Lett. B}\ }\textbf {\bibinfo {volume}
  {381}},\ \bibinfo {pages} {191} (\bibinfo {year} {1996})},\ \Eprint
  {http://arxiv.org/abs/hep-ph/9605230} {arXiv:hep-ph/9605230} \BibitemShut
  {NoStop}%
\bibitem [{\citenamefont {Dias}\ \emph {et~al.}(2003)\citenamefont {Dias},
  \citenamefont {Pleitez},\ and\ \citenamefont {Tonasse}}]{Dias:2002gg}%
  \BibitemOpen
  \bibfield  {author} {\bibinfo {author} {\bibfnamefont {A.~G.}\ \bibnamefont
  {Dias}}, \bibinfo {author} {\bibfnamefont {V.}~\bibnamefont {Pleitez}}, \
  and\ \bibinfo {author} {\bibfnamefont {M.~D.}\ \bibnamefont {Tonasse}},\
  }\href {\doibase 10.1103/PhysRevD.67.095008} {\bibfield  {journal} {\bibinfo
  {journal} {Phys. Rev. D}\ }\textbf {\bibinfo {volume} {67}},\ \bibinfo
  {pages} {095008} (\bibinfo {year} {2003})},\ \Eprint
  {http://arxiv.org/abs/hep-ph/0211107} {arXiv:hep-ph/0211107} \BibitemShut
  {NoStop}%
\bibitem [{\citenamefont {Ferreira}\ \emph {et~al.}(2011)\citenamefont
  {Ferreira}, \citenamefont {Pinheiro}, \citenamefont {Pires},\ and\
  \citenamefont {da~Silva}}]{Ferreira:2011hm}%
  \BibitemOpen
  \bibfield  {author} {\bibinfo {author} {\bibfnamefont {J.~G.}\ \bibnamefont
  {Ferreira}, \bibfnamefont {Jr}}, \bibinfo {author} {\bibfnamefont {P.~R.~D.}\
  \bibnamefont {Pinheiro}}, \bibinfo {author} {\bibfnamefont {C.~A. d.~S.}\
  \bibnamefont {Pires}}, \ and\ \bibinfo {author} {\bibfnamefont {P.~S.~R.}\
  \bibnamefont {da~Silva}},\ }\href {\doibase 10.1103/PhysRevD.84.095019}
  {\bibfield  {journal} {\bibinfo  {journal} {Phys. Rev. D}\ }\textbf {\bibinfo
  {volume} {84}},\ \bibinfo {pages} {095019} (\bibinfo {year} {2011})},\
  \Eprint {http://arxiv.org/abs/1109.0031} {arXiv:1109.0031 [hep-ph]}
  \BibitemShut {NoStop}%
\bibitem [{\citenamefont {Witten}(1982)}]{Witten:1982fp}%
  \BibitemOpen
  \bibfield  {author} {\bibinfo {author} {\bibfnamefont {E.}~\bibnamefont
  {Witten}},\ }\href {\doibase 10.1016/0370-2693(82)90728-6} {\bibfield
  {journal} {\bibinfo  {journal} {Phys. Lett. B}\ }\textbf {\bibinfo {volume}
  {117}},\ \bibinfo {pages} {324} (\bibinfo {year} {1982})}\BibitemShut
  {NoStop}%
\bibitem [{\citenamefont {Li}(1974)}]{Li:1973mq}%
  \BibitemOpen
  \bibfield  {author} {\bibinfo {author} {\bibfnamefont {L.-F.}\ \bibnamefont
  {Li}},\ }\href {\doibase 10.1103/PhysRevD.9.1723} {\bibfield  {journal}
  {\bibinfo  {journal} {Phys. Rev. D}\ }\textbf {\bibinfo {volume} {9}},\
  \bibinfo {pages} {1723} (\bibinfo {year} {1974})}\BibitemShut {NoStop}%
\bibitem [{\citenamefont {Chen}\ \emph {et~al.}(2010)\citenamefont {Chen},
  \citenamefont {Ryttov},\ and\ \citenamefont {Shrock}}]{Chen:2010er}%
  \BibitemOpen
  \bibfield  {author} {\bibinfo {author} {\bibfnamefont {N.}~\bibnamefont
  {Chen}}, \bibinfo {author} {\bibfnamefont {T.~A.}\ \bibnamefont {Ryttov}}, \
  and\ \bibinfo {author} {\bibfnamefont {R.}~\bibnamefont {Shrock}},\ }\href
  {\doibase 10.1103/PhysRevD.82.116006} {\bibfield  {journal} {\bibinfo
  {journal} {Phys. Rev. D}\ }\textbf {\bibinfo {volume} {82}},\ \bibinfo
  {pages} {116006} (\bibinfo {year} {2010})},\ \Eprint
  {http://arxiv.org/abs/1010.3736} {arXiv:1010.3736 [hep-ph]} \BibitemShut
  {NoStop}%
\bibitem [{\citenamefont {Di~Luzio}\ \emph {et~al.}(2015)\citenamefont
  {Di~Luzio}, \citenamefont {Gr\"ober}, \citenamefont {Kamenik},\ and\
  \citenamefont {Nardecchia}}]{DiLuzio:2015oha}%
  \BibitemOpen
  \bibfield  {author} {\bibinfo {author} {\bibfnamefont {L.}~\bibnamefont
  {Di~Luzio}}, \bibinfo {author} {\bibfnamefont {R.}~\bibnamefont {Gr\"ober}},
  \bibinfo {author} {\bibfnamefont {J.~F.}\ \bibnamefont {Kamenik}}, \ and\
  \bibinfo {author} {\bibfnamefont {M.}~\bibnamefont {Nardecchia}},\ }\href
  {\doibase 10.1007/JHEP07(2015)074} {\bibfield  {journal} {\bibinfo  {journal}
  {JHEP}\ }\textbf {\bibinfo {volume} {07}},\ \bibinfo {pages} {074} (\bibinfo
  {year} {2015})},\ \Eprint {http://arxiv.org/abs/1504.00359} {arXiv:1504.00359
  [hep-ph]} \BibitemShut {NoStop}%
\bibitem [{\citenamefont {Di~Luzio}\ \emph
  {et~al.}(2017{\natexlab{b}})\citenamefont {Di~Luzio}, \citenamefont
  {Mescia},\ and\ \citenamefont {Nardi}}]{DiLuzio:2016sbl}%
  \BibitemOpen
  \bibfield  {author} {\bibinfo {author} {\bibfnamefont {L.}~\bibnamefont
  {Di~Luzio}}, \bibinfo {author} {\bibfnamefont {F.}~\bibnamefont {Mescia}}, \
  and\ \bibinfo {author} {\bibfnamefont {E.}~\bibnamefont {Nardi}},\ }\href
  {\doibase 10.1103/PhysRevLett.118.031801} {\bibfield  {journal} {\bibinfo
  {journal} {Phys. Rev. Lett.}\ }\textbf {\bibinfo {volume} {118}},\ \bibinfo
  {pages} {031801} (\bibinfo {year} {2017}{\natexlab{b}})},\ \Eprint
  {http://arxiv.org/abs/1610.07593} {arXiv:1610.07593 [hep-ph]} \BibitemShut
  {NoStop}%
\bibitem [{\citenamefont {Di~Luzio}\ \emph
  {et~al.}(2017{\natexlab{c}})\citenamefont {Di~Luzio}, \citenamefont
  {Mescia},\ and\ \citenamefont {Nardi}}]{DiLuzio:2017pfr}%
  \BibitemOpen
  \bibfield  {author} {\bibinfo {author} {\bibfnamefont {L.}~\bibnamefont
  {Di~Luzio}}, \bibinfo {author} {\bibfnamefont {F.}~\bibnamefont {Mescia}}, \
  and\ \bibinfo {author} {\bibfnamefont {E.}~\bibnamefont {Nardi}},\ }\href
  {\doibase 10.1103/PhysRevD.96.075003} {\bibfield  {journal} {\bibinfo
  {journal} {Phys. Rev. D}\ }\textbf {\bibinfo {volume} {96}},\ \bibinfo
  {pages} {075003} (\bibinfo {year} {2017}{\natexlab{c}})},\ \Eprint
  {http://arxiv.org/abs/1705.05370} {arXiv:1705.05370 [hep-ph]} \BibitemShut
  {NoStop}%
\bibitem [{\citenamefont {Kawasaki}\ \emph
  {et~al.}(2005{\natexlab{a}})\citenamefont {Kawasaki}, \citenamefont {Kohri},\
  and\ \citenamefont {Moroi}}]{Kawasaki:2004qu}%
  \BibitemOpen
  \bibfield  {author} {\bibinfo {author} {\bibfnamefont {M.}~\bibnamefont
  {Kawasaki}}, \bibinfo {author} {\bibfnamefont {K.}~\bibnamefont {Kohri}}, \
  and\ \bibinfo {author} {\bibfnamefont {T.}~\bibnamefont {Moroi}},\ }\href
  {\doibase 10.1103/PhysRevD.71.083502} {\bibfield  {journal} {\bibinfo
  {journal} {Phys. Rev. D}\ }\textbf {\bibinfo {volume} {71}},\ \bibinfo
  {pages} {083502} (\bibinfo {year} {2005}{\natexlab{a}})},\ \Eprint
  {http://arxiv.org/abs/astro-ph/0408426} {arXiv:astro-ph/0408426} \BibitemShut
  {NoStop}%
\bibitem [{\citenamefont {Jedamzik}(2008)}]{Jedamzik:2007qk}%
  \BibitemOpen
  \bibfield  {author} {\bibinfo {author} {\bibfnamefont {K.}~\bibnamefont
  {Jedamzik}},\ }\href {\doibase 10.1088/1475-7516/2008/03/008} {\bibfield
  {journal} {\bibinfo  {journal} {JCAP}\ }\textbf {\bibinfo {volume} {03}},\
  \bibinfo {pages} {008} (\bibinfo {year} {2008})},\ \Eprint
  {http://arxiv.org/abs/0710.5153} {arXiv:0710.5153 [hep-ph]} \BibitemShut
  {NoStop}%
\bibitem [{\citenamefont {Davidson}\ and\ \citenamefont
  {Ibarra}(2002)}]{Davidson:2002qv}%
  \BibitemOpen
  \bibfield  {author} {\bibinfo {author} {\bibfnamefont {S.}~\bibnamefont
  {Davidson}}\ and\ \bibinfo {author} {\bibfnamefont {A.}~\bibnamefont
  {Ibarra}},\ }\href {\doibase 10.1016/S0370-2693(02)01735-5} {\bibfield
  {journal} {\bibinfo  {journal} {Phys. Lett. B}\ }\textbf {\bibinfo {volume}
  {535}},\ \bibinfo {pages} {25} (\bibinfo {year} {2002})},\ \Eprint
  {http://arxiv.org/abs/hep-ph/0202239} {arXiv:hep-ph/0202239} \BibitemShut
  {NoStop}%
\bibitem [{\citenamefont {Buchmuller}\ \emph {et~al.}(2002)\citenamefont
  {Buchmuller}, \citenamefont {Di~Bari},\ and\ \citenamefont
  {Plumacher}}]{Buchmuller:2002rq}%
  \BibitemOpen
  \bibfield  {author} {\bibinfo {author} {\bibfnamefont {W.}~\bibnamefont
  {Buchmuller}}, \bibinfo {author} {\bibfnamefont {P.}~\bibnamefont {Di~Bari}},
  \ and\ \bibinfo {author} {\bibfnamefont {M.}~\bibnamefont {Plumacher}},\
  }\href {\doibase 10.1016/S0550-3213(02)00737-X} {\bibfield  {journal}
  {\bibinfo  {journal} {Nucl. Phys. B}\ }\textbf {\bibinfo {volume} {643}},\
  \bibinfo {pages} {367} (\bibinfo {year} {2002})},\ \bibinfo {note} {[Erratum:
  Nucl.Phys.B 793, 362 (2008)]},\ \Eprint {http://arxiv.org/abs/hep-ph/0205349}
  {arXiv:hep-ph/0205349} \BibitemShut {NoStop}%
\bibitem [{\citenamefont {Ellis}\ and\ \citenamefont
  {Raidal}(2002)}]{Ellis:2002xg}%
  \BibitemOpen
  \bibfield  {author} {\bibinfo {author} {\bibfnamefont {J.~R.}\ \bibnamefont
  {Ellis}}\ and\ \bibinfo {author} {\bibfnamefont {M.}~\bibnamefont {Raidal}},\
  }\href {\doibase 10.1016/S0550-3213(02)00753-8} {\bibfield  {journal}
  {\bibinfo  {journal} {Nucl. Phys. B}\ }\textbf {\bibinfo {volume} {643}},\
  \bibinfo {pages} {229} (\bibinfo {year} {2002})},\ \Eprint
  {http://arxiv.org/abs/hep-ph/0206174} {arXiv:hep-ph/0206174} \BibitemShut
  {NoStop}%
\bibitem [{\citenamefont {'t~Hooft}(1980)}]{tHooft:1979rat}%
  \BibitemOpen
  \bibfield  {author} {\bibinfo {author} {\bibfnamefont {G.}~\bibnamefont
  {'t~Hooft}},\ }\href {\doibase 10.1007/978-1-4684-7571-5_9} {\bibfield
  {journal} {\bibinfo  {journal} {NATO Sci. Ser. B}\ }\textbf {\bibinfo
  {volume} {59}},\ \bibinfo {pages} {135} (\bibinfo {year} {1980})}\BibitemShut
  {NoStop}%
\bibitem [{\citenamefont {Chang}\ \emph {et~al.}(2018)\citenamefont {Chang},
  \citenamefont {Essig},\ and\ \citenamefont {McDermott}}]{Chang:2018rso}%
  \BibitemOpen
  \bibfield  {author} {\bibinfo {author} {\bibfnamefont {J.~H.}\ \bibnamefont
  {Chang}}, \bibinfo {author} {\bibfnamefont {R.}~\bibnamefont {Essig}}, \ and\
  \bibinfo {author} {\bibfnamefont {S.~D.}\ \bibnamefont {McDermott}},\ }\href
  {\doibase 10.1007/JHEP09(2018)051} {\bibfield  {journal} {\bibinfo  {journal}
  {JHEP}\ }\textbf {\bibinfo {volume} {09}},\ \bibinfo {pages} {051} (\bibinfo
  {year} {2018})},\ \Eprint {http://arxiv.org/abs/1803.00993} {arXiv:1803.00993
  [hep-ph]} \BibitemShut {NoStop}%
\bibitem [{\citenamefont {Georgi}\ \emph {et~al.}(1986)\citenamefont {Georgi},
  \citenamefont {Kaplan},\ and\ \citenamefont {Randall}}]{Georgi:1986df}%
  \BibitemOpen
  \bibfield  {author} {\bibinfo {author} {\bibfnamefont {H.}~\bibnamefont
  {Georgi}}, \bibinfo {author} {\bibfnamefont {D.~B.}\ \bibnamefont {Kaplan}},
  \ and\ \bibinfo {author} {\bibfnamefont {L.}~\bibnamefont {Randall}},\ }\href
  {\doibase 10.1016/0370-2693(86)90688-X} {\bibfield  {journal} {\bibinfo
  {journal} {Phys. Lett. B}\ }\textbf {\bibinfo {volume} {169}},\ \bibinfo
  {pages} {73} (\bibinfo {year} {1986})}\BibitemShut {NoStop}%
\bibitem [{\citenamefont {Armengaud}\ \emph {et~al.}(2014)\citenamefont
  {Armengaud} \emph {et~al.}}]{Armengaud:2014gea}%
  \BibitemOpen
  \bibfield  {author} {\bibinfo {author} {\bibfnamefont {E.}~\bibnamefont
  {Armengaud}} \emph {et~al.},\ }\href {\doibase 10.1088/1748-0221/9/05/T05002}
  {\bibfield  {journal} {\bibinfo  {journal} {JINST}\ }\textbf {\bibinfo
  {volume} {9}},\ \bibinfo {pages} {T05002} (\bibinfo {year} {2014})},\ \Eprint
  {http://arxiv.org/abs/1401.3233} {arXiv:1401.3233 [physics.ins-det]}
  \BibitemShut {NoStop}%
\bibitem [{\citenamefont {Armengaud}\ \emph {et~al.}(2019)\citenamefont
  {Armengaud} \emph {et~al.}}]{Armengaud:2019uso}%
  \BibitemOpen
  \bibfield  {author} {\bibinfo {author} {\bibfnamefont {E.}~\bibnamefont
  {Armengaud}} \emph {et~al.} (\bibinfo {collaboration} {IAXO}),\ }\href
  {\doibase 10.1088/1475-7516/2019/06/047} {\bibfield  {journal} {\bibinfo
  {journal} {JCAP}\ }\textbf {\bibinfo {volume} {06}},\ \bibinfo {pages} {047}
  (\bibinfo {year} {2019})},\ \Eprint {http://arxiv.org/abs/1904.09155}
  {arXiv:1904.09155 [hep-ph]} \BibitemShut {NoStop}%
\bibitem [{\citenamefont {Ge}\ \emph {et~al.}(2020)\citenamefont {Ge},
  \citenamefont {Hamaguchi}, \citenamefont {Ichimura}, \citenamefont
  {Ishidoshiro}, \citenamefont {Kanazawa}, \citenamefont {Kishimoto},
  \citenamefont {Nagata},\ and\ \citenamefont {Zheng}}]{Ge:2020zww}%
  \BibitemOpen
  \bibfield  {author} {\bibinfo {author} {\bibfnamefont {S.-F.}\ \bibnamefont
  {Ge}}, \bibinfo {author} {\bibfnamefont {K.}~\bibnamefont {Hamaguchi}},
  \bibinfo {author} {\bibfnamefont {K.}~\bibnamefont {Ichimura}}, \bibinfo
  {author} {\bibfnamefont {K.}~\bibnamefont {Ishidoshiro}}, \bibinfo {author}
  {\bibfnamefont {Y.}~\bibnamefont {Kanazawa}}, \bibinfo {author}
  {\bibfnamefont {Y.}~\bibnamefont {Kishimoto}}, \bibinfo {author}
  {\bibfnamefont {N.}~\bibnamefont {Nagata}}, \ and\ \bibinfo {author}
  {\bibfnamefont {J.}~\bibnamefont {Zheng}},\ }\href {\doibase
  10.1088/1475-7516/2020/11/059} {\bibfield  {journal} {\bibinfo  {journal}
  {JCAP}\ }\textbf {\bibinfo {volume} {11}},\ \bibinfo {pages} {059} (\bibinfo
  {year} {2020})},\ \Eprint {http://arxiv.org/abs/2008.03924} {arXiv:2008.03924
  [hep-ph]} \BibitemShut {NoStop}%
\bibitem [{\citenamefont {Baryakhtar}\ \emph {et~al.}(2018)\citenamefont
  {Baryakhtar}, \citenamefont {Huang},\ and\ \citenamefont
  {Lasenby}}]{Baryakhtar:2018doz}%
  \BibitemOpen
  \bibfield  {author} {\bibinfo {author} {\bibfnamefont {M.}~\bibnamefont
  {Baryakhtar}}, \bibinfo {author} {\bibfnamefont {J.}~\bibnamefont {Huang}}, \
  and\ \bibinfo {author} {\bibfnamefont {R.}~\bibnamefont {Lasenby}},\ }\href
  {\doibase 10.1103/PhysRevD.98.035006} {\bibfield  {journal} {\bibinfo
  {journal} {Phys. Rev. D}\ }\textbf {\bibinfo {volume} {98}},\ \bibinfo
  {pages} {035006} (\bibinfo {year} {2018})},\ \Eprint
  {http://arxiv.org/abs/1803.11455} {arXiv:1803.11455 [hep-ph]} \BibitemShut
  {NoStop}%
\bibitem [{\citenamefont {Lawson}\ \emph {et~al.}(2019)\citenamefont {Lawson},
  \citenamefont {Millar}, \citenamefont {Pancaldi}, \citenamefont
  {Vitagliano},\ and\ \citenamefont {Wilczek}}]{Lawson:2019brd}%
  \BibitemOpen
  \bibfield  {author} {\bibinfo {author} {\bibfnamefont {M.}~\bibnamefont
  {Lawson}}, \bibinfo {author} {\bibfnamefont {A.~J.}\ \bibnamefont {Millar}},
  \bibinfo {author} {\bibfnamefont {M.}~\bibnamefont {Pancaldi}}, \bibinfo
  {author} {\bibfnamefont {E.}~\bibnamefont {Vitagliano}}, \ and\ \bibinfo
  {author} {\bibfnamefont {F.}~\bibnamefont {Wilczek}},\ }\href {\doibase
  10.1103/PhysRevLett.123.141802} {\bibfield  {journal} {\bibinfo  {journal}
  {Phys. Rev. Lett.}\ }\textbf {\bibinfo {volume} {123}},\ \bibinfo {pages}
  {141802} (\bibinfo {year} {2019})},\ \Eprint
  {http://arxiv.org/abs/1904.11872} {arXiv:1904.11872 [hep-ph]} \BibitemShut
  {NoStop}%
\bibitem [{\citenamefont {Di~Luzio}\ \emph {et~al.}(2020)\citenamefont
  {Di~Luzio}, \citenamefont {Giannotti}, \citenamefont {Nardi},\ and\
  \citenamefont {Visinelli}}]{DiLuzio:2020wdo}%
  \BibitemOpen
  \bibfield  {author} {\bibinfo {author} {\bibfnamefont {L.}~\bibnamefont
  {Di~Luzio}}, \bibinfo {author} {\bibfnamefont {M.}~\bibnamefont {Giannotti}},
  \bibinfo {author} {\bibfnamefont {E.}~\bibnamefont {Nardi}}, \ and\ \bibinfo
  {author} {\bibfnamefont {L.}~\bibnamefont {Visinelli}},\ }\href {\doibase
  10.1016/j.physrep.2020.06.002} {\bibfield  {journal} {\bibinfo  {journal}
  {Phys. Rept.}\ }\textbf {\bibinfo {volume} {870}},\ \bibinfo {pages} {1}
  (\bibinfo {year} {2020})},\ \Eprint {http://arxiv.org/abs/2003.01100}
  {arXiv:2003.01100 [hep-ph]} \BibitemShut {NoStop}%
\bibitem [{\citenamefont {Vilenkin}(1981)}]{Vilenkin:1981zs}%
  \BibitemOpen
  \bibfield  {author} {\bibinfo {author} {\bibfnamefont {A.}~\bibnamefont
  {Vilenkin}},\ }\href {\doibase 10.1103/PhysRevD.23.852} {\bibfield  {journal}
  {\bibinfo  {journal} {Phys. Rev. D}\ }\textbf {\bibinfo {volume} {23}},\
  \bibinfo {pages} {852} (\bibinfo {year} {1981})}\BibitemShut {NoStop}%
\bibitem [{\citenamefont {Gelmini}\ \emph {et~al.}(1989)\citenamefont
  {Gelmini}, \citenamefont {Gleiser},\ and\ \citenamefont
  {Kolb}}]{Gelmini:1988sf}%
  \BibitemOpen
  \bibfield  {author} {\bibinfo {author} {\bibfnamefont {G.~B.}\ \bibnamefont
  {Gelmini}}, \bibinfo {author} {\bibfnamefont {M.}~\bibnamefont {Gleiser}}, \
  and\ \bibinfo {author} {\bibfnamefont {E.~W.}\ \bibnamefont {Kolb}},\ }\href
  {\doibase 10.1103/PhysRevD.39.1558} {\bibfield  {journal} {\bibinfo
  {journal} {Phys. Rev. D}\ }\textbf {\bibinfo {volume} {39}},\ \bibinfo
  {pages} {1558} (\bibinfo {year} {1989})}\BibitemShut {NoStop}%
\bibitem [{\citenamefont {Larsson}\ \emph {et~al.}(1997)\citenamefont
  {Larsson}, \citenamefont {Sarkar},\ and\ \citenamefont
  {White}}]{Larsson:1996sp}%
  \BibitemOpen
  \bibfield  {author} {\bibinfo {author} {\bibfnamefont {S.~E.}\ \bibnamefont
  {Larsson}}, \bibinfo {author} {\bibfnamefont {S.}~\bibnamefont {Sarkar}}, \
  and\ \bibinfo {author} {\bibfnamefont {P.~L.}\ \bibnamefont {White}},\ }\href
  {\doibase 10.1103/PhysRevD.55.5129} {\bibfield  {journal} {\bibinfo
  {journal} {Phys. Rev. D}\ }\textbf {\bibinfo {volume} {55}},\ \bibinfo
  {pages} {5129} (\bibinfo {year} {1997})},\ \Eprint
  {http://arxiv.org/abs/hep-ph/9608319} {arXiv:hep-ph/9608319} \BibitemShut
  {NoStop}%
\bibitem [{\citenamefont {Kawasaki}\ \emph
  {et~al.}(2005{\natexlab{b}})\citenamefont {Kawasaki}, \citenamefont {Kohri},\
  and\ \citenamefont {Moroi}}]{Kawasaki:2004yh}%
  \BibitemOpen
  \bibfield  {author} {\bibinfo {author} {\bibfnamefont {M.}~\bibnamefont
  {Kawasaki}}, \bibinfo {author} {\bibfnamefont {K.}~\bibnamefont {Kohri}}, \
  and\ \bibinfo {author} {\bibfnamefont {T.}~\bibnamefont {Moroi}},\ }\href
  {\doibase 10.1016/j.physletb.2005.08.045} {\bibfield  {journal} {\bibinfo
  {journal} {Phys. Lett. B}\ }\textbf {\bibinfo {volume} {625}},\ \bibinfo
  {pages} {7} (\bibinfo {year} {2005}{\natexlab{b}})},\ \Eprint
  {http://arxiv.org/abs/astro-ph/0402490} {arXiv:astro-ph/0402490} \BibitemShut
  {NoStop}%
\bibitem [{\citenamefont {Saikawa}(2017)}]{Saikawa:2017hiv}%
  \BibitemOpen
  \bibfield  {author} {\bibinfo {author} {\bibfnamefont {K.}~\bibnamefont
  {Saikawa}},\ }\href {\doibase 10.3390/universe3020040} {\bibfield  {journal}
  {\bibinfo  {journal} {Universe}\ }\textbf {\bibinfo {volume} {3}},\ \bibinfo
  {pages} {40} (\bibinfo {year} {2017})},\ \Eprint
  {http://arxiv.org/abs/1703.02576} {arXiv:1703.02576 [hep-ph]} \BibitemShut
  {NoStop}%
\bibitem [{\citenamefont {Sikivie}(1982)}]{Sikivie:1982qv}%
  \BibitemOpen
  \bibfield  {author} {\bibinfo {author} {\bibfnamefont {P.}~\bibnamefont
  {Sikivie}},\ }\href {\doibase 10.1103/PhysRevLett.48.1156} {\bibfield
  {journal} {\bibinfo  {journal} {Phys. Rev. Lett.}\ }\textbf {\bibinfo
  {volume} {48}},\ \bibinfo {pages} {1156} (\bibinfo {year}
  {1982})}\BibitemShut {NoStop}%
\bibitem [{\citenamefont {Huang}\ and\ \citenamefont
  {Sikivie}(1985)}]{Huang:1985tt}%
  \BibitemOpen
  \bibfield  {author} {\bibinfo {author} {\bibfnamefont {M.~C.}\ \bibnamefont
  {Huang}}\ and\ \bibinfo {author} {\bibfnamefont {P.}~\bibnamefont
  {Sikivie}},\ }\href {\doibase 10.1103/PhysRevD.32.1560} {\bibfield  {journal}
  {\bibinfo  {journal} {Phys. Rev. D}\ }\textbf {\bibinfo {volume} {32}},\
  \bibinfo {pages} {1560} (\bibinfo {year} {1985})}\BibitemShut {NoStop}%
\bibitem [{\citenamefont {Hiramatsu}\ \emph {et~al.}(2013)\citenamefont
  {Hiramatsu}, \citenamefont {Kawasaki}, \citenamefont {Saikawa},\ and\
  \citenamefont {Sekiguchi}}]{Hiramatsu:2012sc}%
  \BibitemOpen
  \bibfield  {author} {\bibinfo {author} {\bibfnamefont {T.}~\bibnamefont
  {Hiramatsu}}, \bibinfo {author} {\bibfnamefont {M.}~\bibnamefont {Kawasaki}},
  \bibinfo {author} {\bibfnamefont {K.}~\bibnamefont {Saikawa}}, \ and\
  \bibinfo {author} {\bibfnamefont {T.}~\bibnamefont {Sekiguchi}},\ }\href
  {\doibase 10.1088/1475-7516/2013/01/001} {\bibfield  {journal} {\bibinfo
  {journal} {JCAP}\ }\textbf {\bibinfo {volume} {01}},\ \bibinfo {pages} {001}
  (\bibinfo {year} {2013})},\ \Eprint {http://arxiv.org/abs/1207.3166}
  {arXiv:1207.3166 [hep-ph]} \BibitemShut {NoStop}%
\bibitem [{\citenamefont {Tanabashi}\ \emph {et~al.}(2018)\citenamefont
  {Tanabashi} \emph {et~al.}}]{Tanabashi:2018oca}%
  \BibitemOpen
  \bibfield  {author} {\bibinfo {author} {\bibfnamefont {M.}~\bibnamefont
  {Tanabashi}} \emph {et~al.} (\bibinfo {collaboration} {Particle Data
  Group}),\ }\href {\doibase 10.1103/PhysRevD.98.030001} {\bibfield  {journal}
  {\bibinfo  {journal} {Phys. Rev. D}\ }\textbf {\bibinfo {volume} {98}},\
  \bibinfo {pages} {030001} (\bibinfo {year} {2018})}\BibitemShut {NoStop}%
\bibitem [{\citenamefont {Weinberg}(1980)}]{Weinberg:1980wa}%
  \BibitemOpen
  \bibfield  {author} {\bibinfo {author} {\bibfnamefont {S.}~\bibnamefont
  {Weinberg}},\ }\href {\doibase 10.1016/0370-2693(80)90660-7} {\bibfield
  {journal} {\bibinfo  {journal} {Phys. Lett. B}\ }\textbf {\bibinfo {volume}
  {91}},\ \bibinfo {pages} {51} (\bibinfo {year} {1980})}\BibitemShut {NoStop}%
\bibitem [{\citenamefont {Hall}(1981)}]{Hall:1980kf}%
  \BibitemOpen
  \bibfield  {author} {\bibinfo {author} {\bibfnamefont {L.~J.}\ \bibnamefont
  {Hall}},\ }\href {\doibase 10.1016/0550-3213(81)90498-3} {\bibfield
  {journal} {\bibinfo  {journal} {Nucl. Phys. B}\ }\textbf {\bibinfo {volume}
  {178}},\ \bibinfo {pages} {75} (\bibinfo {year} {1981})}\BibitemShut
  {NoStop}%
\bibitem [{\citenamefont {Chakrabortty}\ \emph {et~al.}(2019)\citenamefont
  {Chakrabortty}, \citenamefont {Maji},\ and\ \citenamefont
  {King}}]{Chakrabortty:2019fov}%
  \BibitemOpen
  \bibfield  {author} {\bibinfo {author} {\bibfnamefont {J.}~\bibnamefont
  {Chakrabortty}}, \bibinfo {author} {\bibfnamefont {R.}~\bibnamefont {Maji}},
  \ and\ \bibinfo {author} {\bibfnamefont {S.~F.}\ \bibnamefont {King}},\
  }\href {\doibase 10.1103/PhysRevD.99.095008} {\bibfield  {journal} {\bibinfo
  {journal} {Phys. Rev. D}\ }\textbf {\bibinfo {volume} {99}},\ \bibinfo
  {pages} {095008} (\bibinfo {year} {2019})},\ \Eprint
  {http://arxiv.org/abs/1901.05867} {arXiv:1901.05867 [hep-ph]} \BibitemShut
  {NoStop}%
\bibitem [{\citenamefont {Meloni}\ \emph {et~al.}(2020)\citenamefont {Meloni},
  \citenamefont {Ohlsson},\ and\ \citenamefont {Pernow}}]{Meloni:2019jcf}%
  \BibitemOpen
  \bibfield  {author} {\bibinfo {author} {\bibfnamefont {D.}~\bibnamefont
  {Meloni}}, \bibinfo {author} {\bibfnamefont {T.}~\bibnamefont {Ohlsson}}, \
  and\ \bibinfo {author} {\bibfnamefont {M.}~\bibnamefont {Pernow}},\ }\href
  {\doibase 10.1140/epjc/s10052-020-8308-9} {\bibfield  {journal} {\bibinfo
  {journal} {Eur. Phys. J. C}\ }\textbf {\bibinfo {volume} {80}},\ \bibinfo
  {pages} {840} (\bibinfo {year} {2020})},\ \Eprint
  {http://arxiv.org/abs/1911.11411} {arXiv:1911.11411 [hep-ph]} \BibitemShut
  {NoStop}%
\bibitem [{\citenamefont {Dash}\ \emph {et~al.}(2021)\citenamefont {Dash},
  \citenamefont {Mishra}, \citenamefont {Patra},\ and\ \citenamefont
  {Sahu}}]{Dash:2020jlc}%
  \BibitemOpen
  \bibfield  {author} {\bibinfo {author} {\bibfnamefont {C.}~\bibnamefont
  {Dash}}, \bibinfo {author} {\bibfnamefont {S.}~\bibnamefont {Mishra}},
  \bibinfo {author} {\bibfnamefont {S.}~\bibnamefont {Patra}}, \ and\ \bibinfo
  {author} {\bibfnamefont {P.}~\bibnamefont {Sahu}},\ }\href {\doibase
  10.1016/j.nuclphysb.2020.115239} {\bibfield  {journal} {\bibinfo  {journal}
  {Nucl. Phys. B}\ }\textbf {\bibinfo {volume} {962}},\ \bibinfo {pages}
  {115239} (\bibinfo {year} {2021})},\ \Eprint
  {http://arxiv.org/abs/2004.14188} {arXiv:2004.14188 [hep-ph]} \BibitemShut
  {NoStop}%
\bibitem [{\citenamefont {Ohlsson}\ \emph {et~al.}(2020)\citenamefont
  {Ohlsson}, \citenamefont {Pernow},\ and\ \citenamefont
  {S\"onnerlind}}]{Ohlsson:2020rjc}%
  \BibitemOpen
  \bibfield  {author} {\bibinfo {author} {\bibfnamefont {T.}~\bibnamefont
  {Ohlsson}}, \bibinfo {author} {\bibfnamefont {M.}~\bibnamefont {Pernow}}, \
  and\ \bibinfo {author} {\bibfnamefont {E.}~\bibnamefont {S\"onnerlind}},\
  }\href {\doibase 10.1140/epjc/s10052-020-08679-0} {\bibfield  {journal}
  {\bibinfo  {journal} {Eur. Phys. J. C}\ }\textbf {\bibinfo {volume} {80}},\
  \bibinfo {pages} {1089} (\bibinfo {year} {2020})},\ \Eprint
  {http://arxiv.org/abs/2006.13936} {arXiv:2006.13936 [hep-ph]} \BibitemShut
  {NoStop}%
\bibitem [{\citenamefont {King}\ \emph {et~al.}(2021)\citenamefont {King},
  \citenamefont {Pascoli}, \citenamefont {Turner},\ and\ \citenamefont
  {Zhou}}]{King:2021gmj}%
  \BibitemOpen
  \bibfield  {author} {\bibinfo {author} {\bibfnamefont {S.~F.}\ \bibnamefont
  {King}}, \bibinfo {author} {\bibfnamefont {S.}~\bibnamefont {Pascoli}},
  \bibinfo {author} {\bibfnamefont {J.}~\bibnamefont {Turner}}, \ and\ \bibinfo
  {author} {\bibfnamefont {Y.-L.}\ \bibnamefont {Zhou}},\ }\href {\doibase
  10.1007/JHEP10(2021)225} {\bibfield  {journal} {\bibinfo  {journal} {JHEP}\
  }\textbf {\bibinfo {volume} {10}},\ \bibinfo {pages} {225} (\bibinfo {year}
  {2021})},\ \Eprint {http://arxiv.org/abs/2106.15634} {arXiv:2106.15634
  [hep-ph]} \BibitemShut {NoStop}%
\bibitem [{\citenamefont {Abe}\ \emph {et~al.}(2017)\citenamefont {Abe} \emph
  {et~al.}}]{Miura:2016krn}%
  \BibitemOpen
  \bibfield  {author} {\bibinfo {author} {\bibfnamefont {K.}~\bibnamefont
  {Abe}} \emph {et~al.} (\bibinfo {collaboration} {Super-Kamiokande}),\ }\href
  {\doibase 10.1103/PhysRevD.95.012004} {\bibfield  {journal} {\bibinfo
  {journal} {Phys. Rev. D}\ }\textbf {\bibinfo {volume} {95}},\ \bibinfo
  {pages} {012004} (\bibinfo {year} {2017})},\ \Eprint
  {http://arxiv.org/abs/1610.03597} {arXiv:1610.03597 [hep-ex]} \BibitemShut
  {NoStop}%
\bibitem [{\citenamefont {Takenaka}\ \emph {et~al.}(2020)\citenamefont
  {Takenaka} \emph {et~al.}}]{Takenaka:2020vqy}%
  \BibitemOpen
  \bibfield  {author} {\bibinfo {author} {\bibfnamefont {A.}~\bibnamefont
  {Takenaka}} \emph {et~al.} (\bibinfo {collaboration} {Super-Kamiokande}),\
  }\href {\doibase 10.1103/PhysRevD.102.112011} {\bibfield  {journal} {\bibinfo
   {journal} {Phys. Rev. D}\ }\textbf {\bibinfo {volume} {102}},\ \bibinfo
  {pages} {112011} (\bibinfo {year} {2020})},\ \Eprint
  {http://arxiv.org/abs/2010.16098} {arXiv:2010.16098 [hep-ex]} \BibitemShut
  {NoStop}%
\bibitem [{\citenamefont {Georgi}(1979)}]{Georgi:1979md}%
  \BibitemOpen
  \bibfield  {author} {\bibinfo {author} {\bibfnamefont {H.}~\bibnamefont
  {Georgi}},\ }\href {\doibase 10.1016/0550-3213(79)90497-8} {\bibfield
  {journal} {\bibinfo  {journal} {Nucl. Phys. B}\ }\textbf {\bibinfo {volume}
  {156}},\ \bibinfo {pages} {126} (\bibinfo {year} {1979})}\BibitemShut
  {NoStop}%
\bibitem [{\citenamefont {Frampton}(1979)}]{Frampton:1979cw}%
  \BibitemOpen
  \bibfield  {author} {\bibinfo {author} {\bibfnamefont {P.~H.}\ \bibnamefont
  {Frampton}},\ }\href {\doibase 10.1016/0370-2693(79)90472-6} {\bibfield
  {journal} {\bibinfo  {journal} {Phys. Lett. B}\ }\textbf {\bibinfo {volume}
  {88}},\ \bibinfo {pages} {299} (\bibinfo {year} {1979})}\BibitemShut
  {NoStop}%
\bibitem [{\citenamefont {Frampton}(1980)}]{Frampton:1979tj}%
  \BibitemOpen
  \bibfield  {author} {\bibinfo {author} {\bibfnamefont {P.~H.}\ \bibnamefont
  {Frampton}},\ }\href {\doibase 10.1016/0370-2693(80)90140-9} {\bibfield
  {journal} {\bibinfo  {journal} {Phys. Lett. B}\ }\textbf {\bibinfo {volume}
  {89}},\ \bibinfo {pages} {352} (\bibinfo {year} {1980})}\BibitemShut
  {NoStop}%
\bibitem [{\citenamefont {Frampton}\ and\ \citenamefont
  {Nandi}(1979)}]{Frampton:1979fd}%
  \BibitemOpen
  \bibfield  {author} {\bibinfo {author} {\bibfnamefont {P.}~\bibnamefont
  {Frampton}}\ and\ \bibinfo {author} {\bibfnamefont {S.}~\bibnamefont
  {Nandi}},\ }\href {\doibase 10.1103/PhysRevLett.43.1460} {\bibfield
  {journal} {\bibinfo  {journal} {Phys. Rev. Lett.}\ }\textbf {\bibinfo
  {volume} {43}},\ \bibinfo {pages} {1460} (\bibinfo {year}
  {1979})}\BibitemShut {NoStop}%
\end{thebibliography}

%

\end{document}